\documentclass[%
 reprint,
 amsmath,amssymb,
 aps,
]{revtex4-2}

\usepackage{graphicx}
\usepackage{dcolumn}
\usepackage{bm}
\usepackage{hyperref}
\usepackage{xcolor}
\usepackage[mathlines]{lineno}

\usepackage[normalem]{ulem}

\usepackage{amsmath,amssymb}

\newcommand{\floor}[1]{\left\lfloor #1 \right\rfloor}

\begin{document}

\preprint{APS/123-QED}

\title{Dynamics of a Bottom-Heavy Janus Particle Near a Wall Under Shear Flow}

\author{Zohreh Jalilvand}
\affiliation{Department of Chemical Engineering, City College of New York (CCNY), City University of New York (CUNY), 140th Street {\&} Convent Avenue, New York, New York 10031, United States
}%

\author{Daniele Notarmuzi}%
\email{daniele.notarmuzi@tuwien.ac.at}
\affiliation{
Institute for Theoretical Physics, Technische Universit{\"a}t Wien,
Wiedner Hauptstraße 8-10, A-1040, Vienna, Austria
}%

\author{Ubaldo M. C{\'o}rdova-Figueroa}
\affiliation{
Department of Chemical Engineering, University of Puerto Rico-Mayag{\"u}ez, Mayag{\"u}ez, PR 00681, USA.
}

\author{Emanuela Bianchi}
\affiliation{
  $\,$ Institute for Theoretical Physics, Technische Universit{\"a}t Wien, Wiedner Hauptstraße 8-10, A-1040, Vienna, Austria \\
   and CNR-ISC, Uos Sapienza, Piazzale A. Moro 2, 00185 Roma, Italy
}%

\author{Ilona Kretzschmar}
\email{kretzschmar@ccny.cuny.edu}
\affiliation{ Department of Chemical Engineering, City College of New York (CCNY), City University of New York (CUNY), 140th Street {\&} Convent Avenue, New York, New York 10031, United States
}%

\date{\today}

\begin{abstract}
In this study, Brownian Dynamics simulations are implemented to investigate the motion of a bottom-heavy Janus particle near a wall under varying shear flow conditions and at small Péclet ($Pe_p$) numbers. The stochastic motion of the Janus particle impacted by surface forces is described using a set of coupled Langevin equations that takes into account the Janus particle orientation. Interactions arising from surface potentials are found to depend on the separation distance between the Janus particle and the wall, the properties of the surfaces involved, and the thickness of the Janus particle cap. When shear flow is introduced in the system, the dynamical behavior of the Janus particle is also governed by the strain rate. Furthermore, the effect of friction on the dynamical behavior of the Janus particle under shear flow is investigated and reveals that the rotational motion of the Janus particle slows down slightly when the particle is close to the surface. In summary, we demonstrate the ability to utilize Brownian Dynamics simulations to capture the rich dynamical behavior of a bottom-heavy Janus particle near a wall and under a range of shear flow conditions, cap thicknesses, and surface charges.
\end{abstract}

\maketitle

\section{Introduction}
Besides their ubiquity as amphiphilic colloids in nature, Janus colloids have a wide variety of applications in areas such as biomedicine,\cite{SU2019100033,Jeevanandam2023,pharmaceutics15020423} energy storage,\cite{ENAYATIGERDROODBAR2023117831} catalysis,\cite{marc.202300280} and environmental science.\cite{Marschelke2020}
In many of these anticipated applications Janus particles will encounter various types of boundaries such as fluid/fluid interfaces\cite{nano11020374}  and solid walls of blood vessels, channels, and porous structures. As such, understanding and predicting individual particle-wall interactions is important. For example, particle-wall interactions can determine drug delivery rates,\cite{WOS:000442614000018} guide particle swimming,\cite{WOS:000367579100005,WOS:000371019700016,xiao2018review} result in scale formation in pipes,\cite{particledepflow} or change the flow in porous media.\cite{10.1115/1.4051215}

There is a long scientific history regarding the description of particle-wall interactions.\cite{DERJAGUIN199330,alma991011725579703276,10.1063/1.1740347,Israelachvili_1973,RevModPhys.88.045006, doi:10.1021/acs.langmuir.1c01492} Most of the literature addresses spheres or ellipsoids near boundaries with a few studies focusing on spherical passive Janus particles discussed below.
Experimentally, three techniques, Atomic Force Microscopy (AFM), Surface Force Apparatus (SFA), and Total Internal Refection Microscopy (TIRM), have been used to measure particle-wall interactions.
AFM and SFA require the immobilization of the probe particle,\cite{THWALA2023131315, doi:10.1021/acs.langmuir.4c03519, Israelachvili_2010} whereas TIRM does not.
TIRM\cite{PRIEVE199993} is an experimental light scattering technique that indirectly measures particle position above a surface as a function of time resulting in a probability density function (PDF), $p(h)$, in which $h$ is the height of the particle surface from the wall. It then infers the  particle-wall potential, $U(h)$, from analysis of the PDF.
The particle-wall interaction potential and scattering morphology become more complex functions when Janus particles are involved owing to their non-uniform surface charge and potentially heavier cap that influence particle height and orientation in a complex manner.
In a recent perspective article, Yan and Wirth\cite{10.1063/5.0089206} introduce Scattering Morphology Resolved TIRM (SMR-TIRM) and demonstrated that assessment of anisotropic particle interactions near boundaries with SMR-TIRM is more accurate when combined with Brownian Dynamics (BD) simulations.
Using BD simulations, Rashidi et al.\cite{doi:10.1063/1.4994843, PhysRevE.101.042606} explored the dynamics of a polystyrene Janus particle with a non-uniform surface potential with and without a gold cap in the vicinity of a negatively charged wall and successfully derived the 3D potential energy surface of the particle that can be used in the interpretation of TIRM measurements. More recently, the focus has been on understanding the behavior of Janus particles in channels under shear flow, which further complicates the interpretation of TIRM data. Therefore, the development of robust theoretical models and their numerical analysis are a central tool to fully capture the physics of Janus particles, as models are needed to interpret experimental results and, at the same time,
might predict novel behaviors to be in turn verified experimentally.

A number of simulation studies have investigated the dynamic behavior of Janus particle suspensions in channels and near solid boundaries with focus on particle-particle and particle-wall interactions under shear flow.
Using BD simulations, Mohammadi et al.\cite{mohammadi2015brownian} investigated the binding kinetics of Janus particles under shear flow and determined the coagulation rate of a particle pair.
Kobayashi et al.\cite{doi:10.1021/acs.langmuir.6b02694} studied the rheology of Janus nanoparticle suspensions in nanotubes of varying surface chemistry with dissipative particle dynamics and found that flow properties strongly depended on the structure of the colloid phase and the wall chemistry.
Nikoubashman et al.\cite{bianchi2015self,C5SM90068A} employed hybrid molecular dynamics simulations to investigate the shear-induced break up of Janus particle micelles within a slit-like channel under shear flow and found that the shear flow can lead to enhancement of micelles and formation of bigger aggregates some of which show enhanced stability owing to their cluster symmetry. In a follow up study,\cite{C6SM00766J} they found that increasing particle-wall interaction causes adhesion and cluster formation on the channel walls.
DeLaCruz-Araujo et al.\cite{C6SM00183A} showed using BD simulations that exposure of micellar, vesicular, worm-like and lamellar Janus particle aggregates to shear flow induces rearrangement, deformation, and break-up of the aggregates that is more easily controlled by the Peclet number ($Pe_p$) than the interaction potential. In a second study,\cite{delacruz2018shear} they investigated the switch between parallel and perpendicular alignment of lamellar Janus particle aggregates by shear flow and developed a simple scaling argument based on the torque balance on a single Janus particle pair.

Other studies have focused on a single passive Janus particle near a wall. For example, Ramachandran et al.
\cite{ramachandran2009dynamics} used finite element method to study the dynamics and rheology of a dilute suspension of slip-stick Janus particles in creeping flow.
They showed that depending on the ratio of the slip length to the particle’s radius the rotational motion of the particle varies. In addition, the translational motion of the particle is predicted to be impacted by the initial orientation of the particle.
Furthermore, the suspension exhibits non-Newtonian behavior for a specific volume fraction of slip-stick Janus spheres.
Troufa et al.\cite{trofa2019numerical} numerically simulated the dynamics of a slip-stick Janus sphere within a Newtonian fluid confined in a cylindrical micro-channel under Poiseuille flow and found two behavior regimes where the particle either shows periodic oscillations near the wall or migration towards the channel center if near the centerline.

Most recent studies have primarily focused on the steering of active particles near walls,\cite{doi:10.1021/acs.langmuir.0c01924,WOS:000691589900001,PhysRevFluids.9.014202,D4SM00848K,D4SM00733F} though open questions regarding the interactions of passive Janus particles with walls under shear flow persist.
Understanding how the interplay of the particle's and substrate's properties influences the particle’s height, rotation speed, and the frequency with which its surface interacts with the substrate is essential for advancing applications in self-assembly, catalysis, drug delivery, and biosensing.

Here, a BD model is introduced that captures the interaction of a bottom-heavy Janus particle exhibiting varying surface charge with a charged wall with and without shear flow. The BD model is used to fully capture the non-trivial interplay between surface properties of the particle, i.e., surface potential and cap thickness, and the nearby wall in the presence of shear flow, where surface forces and hydrodynamic interactions govern the dynamics and determine the particle height.

\section{Simulation Model}

\begin{figure}[h]
\centering
\includegraphics[width=9cm]{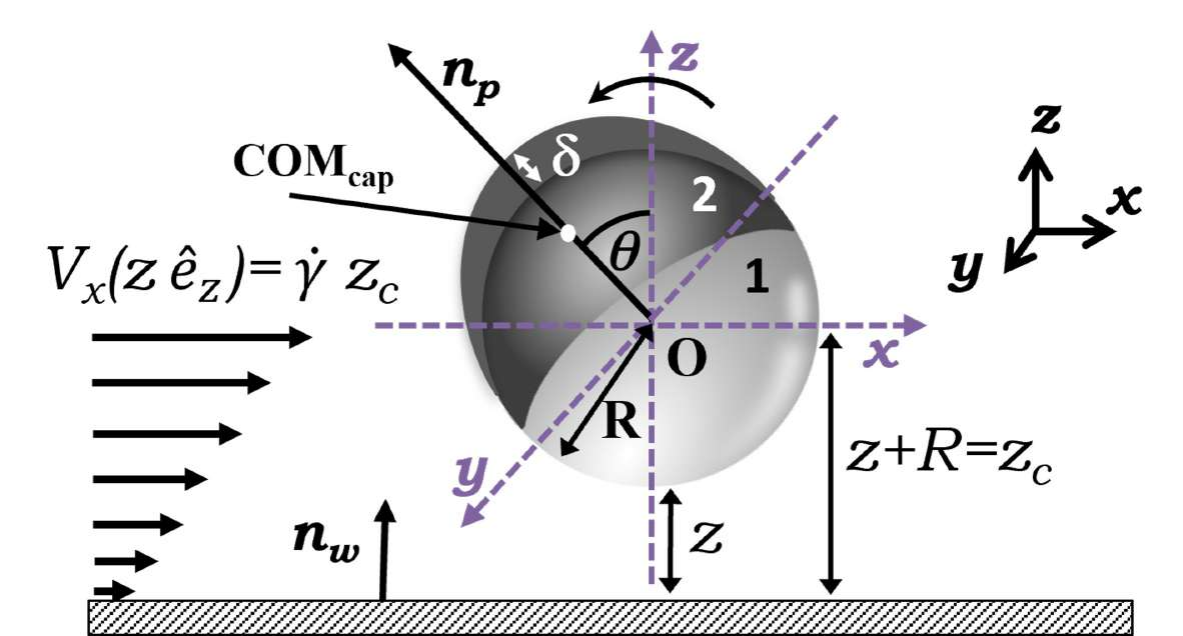}
 \caption{ Schematic representation of a bottom-heavy Janus particle capped on face 2 with a metal of thickness $\delta$ and a cap center of mass COM$_{\textsubscript{cap}}$, bounded with a wall $w$ and subject to shear flow of strain rate $\dot{\gamma}$. Other relevant parameters are also depicted (see text).
 }
\label{fig1}
\end{figure}

A hard sphere of radius $R$ with two different hemispheres
(1,2) serves as a model particle for a system in which each face is uniformly, but independently charged, creating the Janus particle (Fig. \ref{fig1}). The Janus particle is suspended in an aqueous solution of ionic strength $I$, which throughout this study is considered to be deionized (DI) water (Millipore, resistivity $18.2$ M${\Omega}$cm and viscosity $\mu=1e^{-3} Pa\,s$ at $25$ °C, $I=1e^{-6} \,M$). The Janus particle is bounded by a wall, i.e., a substrate made of SiO\textsubscript{2}, which is uniformly charged. The normal to the particle ($\pmb n_p$), that passes through the center of the particle and through the cap's center of mass (COM\textsubscript{cap}) is employed to characterize the rotational behavior of the particle, shown in Figure \ref{fig1}. Finally, the Janus particle is subjected to a shear flow with varying strain rates $\dot{\gamma}$.

Anisotropic interactions between the Janus particle and the wall are modeled by utilizing the total interaction energy, $U_{tot}$, as the sum of electrostatic double layer (EDL) interactions and the gravitational interaction. Hence, $U_{tot}$ is given by equation \ref{eq.1}: \cite{doi:10.1021/acs.langmuir.5b03611,CARNIE1993260}
 \begin{equation}
         U_{tot}(z,\theta, \psi_1,\psi_2,\psi_w)= U_{EDL}(z,\theta,\psi_1,\psi_2,\psi_w)+U_{gravity}(z)\,, \label{eq.1}
\end{equation}
where $z$ denotes the height of the Janus particle, i.e., the distance of the Janus particle from the wall, $\theta$ is defined as the angle between the normal to the particle (i.e., unit vector $\pmb n_p $) and the normal to the wall (i.e., unit vector $\pmb n_w $), and $\psi_1$, $\psi_2$, and $\psi_w$ are the surface potentials of the two faces and the wall, respectively. The contribution from the EDL interaction between a charged particle and a charged wall is given by equation \ref{eq.2}:
\begin{equation}
  \begin{split}
       U_{EDL}(z, \psi_p,\psi_w) & = 2\pi R \varepsilon_0(\psi_p\psi_w \ln{\frac{1+e^{-\kappa z}}{1-e^{-\kappa z}}} - \\
         & \frac{(\psi^2_p+\psi^2_w)}{2}\ln(1-e^{-2\kappa z})) \\
         & :=\frac{1}{\beta}Aw_{EDL}(\bar{z},\phi_p,\phi_w)\,, \label{eq.2}
  \end{split}
\end{equation}
where $A$ is introduced as a dimensionless parameter, equation \ref{eq.3a}, and $w_{EDL}(\bar{z},\phi_p,\phi_w)$ as the dimensionless EDL interaction, equation \ref{eq.3b}:
\begin{subequations}
\begin{eqnarray}
A:= \pi R \varepsilon_0 \psi^2_0 \beta, \label{eq.3a}
 \\
 \begin{split}
	w_{EDL}(\bar{z},\phi_p,\phi_w ) & = 2\phi_p\phi_w \ln {\frac{1+e^{-{\bar{z}}}}{1-e^{-{\bar{z}}}}}-\\
	& (\phi^2_p+\phi^2_w) \ln({1-e^{-2 \bar{z}}})\,,  \label{eq.3b}
 \end{split}
\end{eqnarray}
\end{subequations}
with $\bar{z}=\kappa z$ and the height of the particle $z$ non-dimensionalized by measuring the length scale in units of Debye screening length, $\kappa^{-1}$. In equation \ref{eq.3a}, $\psi_0=1/ q\beta$ is the thermal electrostatic potential used to non-dimensionalize the wall and surface charges as $\phi_i =  \psi_i/\psi_0$ with $i = w$ and $p$, respectively, where $q$ denotes the electron charge (in absolute value) and $\beta=1/ (k_bT)$, where $k_bT$ is the thermal energy. Note the same notation is used in the ESI Section S1,\dag\, which provides additional information on the derivation.

The Janus particle assumes various configurations in proximity to the wall and each hemisphere of the Janus particle contributes its own particle-wall EDL interaction to $U_{EDL}$. The anisotropic interaction is a function of the Janus particle orientation with respect to the wall as specified by $\theta$ (see Fig. \ref{fig1}). The orientation of the cap is incorporated into the model using the orientation-dependent factors $g_{\pm}(\theta) = \frac {(1 \pm \cos \theta)}{2}$ yielding $U_{tot}$ for a Janus particle interacting with a wall, equation \ref{eq.4}:
\begin{equation}
  \begin{split}
         U_{tot}(z,\theta ; \psi_1,\psi_2,\psi_w) & = \frac{A}{\beta} [g_+(\theta)w_{EDL}(\bar{z};\phi_1,\phi_w)+ \\
         & g_-(\theta)w_{EDL}(\bar{z};\phi_2,\phi_w)] \\
         & + U_{gravity}\,.
  \end{split}
  \label{eq.4}
\end{equation}

The governing  equations that describe both the translational and rotational motion of the Janus particle near the wall, known as Langevin equations, are:
\begin{itemize}
\item Translation of particle center, O, along x direction:
\begin{equation}
	\gamma \left(\frac{dx}{dt}-f_xV_x(z_{c})\right)=\ell_x \eta_x(t) \label{eq.5}
\end{equation}
\end{itemize}
\begin{itemize}
\item Translation along z direction:
\begin{equation}
	\gamma \frac{dz}{dt}=F_{EDL}(z,\theta ; \psi_1,\psi_2,\psi_w)+F_g+\ell_z \eta_z(t) \label{eq.6}
\end{equation}
\end{itemize}
\begin{itemize}
\item[$\bullet$] Rotation around $y$ axis (passing through particle center, O):
\begin{equation}
	\xi \left(\frac{d \theta}{dt}-f_{\theta}\frac{1}{2}\dot{\gamma}\right)=N_{EDL}(z,\theta ; \psi_1,\psi_2,\psi_w)+N_g+E \eta_r(t) \label{eq.7} \, .
\end{equation}
\end{itemize}
In the above equations, $\gamma=6\pi\mu R$ and $\xi=8\pi\mu R^3$ are the translational and rotational drag coefficients, respectively, and $f_x$ and $f_{\theta}$ are Goldman et al.'s \cite{goldman1967slow} dimensionless frictional coefficients for translational and rotational motion. Note that expressions for $\gamma$ and $\xi$ represent a perfect sphere and that the contribution from the cap to the particle shape is neglected here. $V_x(z_c)=\dot{\gamma}\, z_c$ denotes the velocity of the external shear flow that is being exerted on the position of the center of the Janus particle, O (see Fig.~\ref{fig1}), by the externally applied strain rate, $\dot{\gamma}$. $\ell_x \eta_z(t)$ and $\ell_z \eta_z(t)$ are the Brownian forces in the $x$ and $z$ directions, respectively, while $E \eta_r(t)$ is the rotational Brownian force. $F_g= m^*g$ and $N_g=-\Xi^* sin(\theta)$ are the gravitational force and torque, respectively, where $m^*$ and $\Xi^*$ correspond to the effective mass of the particle and effective weight of the cap, respectively. Note that in absence of shear flow, $\dot{\gamma}=0$, one clearly has $V_x=0$.
Finally, $F_{EDL}$ and $N_{EDL}$ are the contributions from the electric double layer interaction and defined as:
\begin{gather}
	F_{EDL}(z,\theta ; \psi_1,\psi_2,\psi_w)=-\frac{\partial U_{EDL}}{\partial z}\,, \label{eq.8}\\
	N_{EDL}(z,\theta ; \psi_1,\psi_2,\psi_w)=-\frac{\partial U_{EDL}}{\partial \theta}\,. \label{eq.9}
\end{gather}
The conservative wall-particle interaction forces are computed through the wall-particle interaction energy as $F_{tot}=-\nabla U_{tot}$, which includes the electrostatic double layer interaction and gravitational potential energy. Equations~\ref{eq.5} -~\ref{eq.9} describe the most general system under analysis in this work and is referred to as Model IV.
In particular, they account for different charges of the two sides
of the Janus particle, an asymmetry in the mass distribution as implied by the cap of
thickness $\delta$, the torque induced by the shear
flow, and the hydrodynamic friction. Note that charge imbalance and mass asymmetry induce a force, $F_{EDL}$ and $F_g$, respectively, as well as a torque, $N_{EDL}$ and $N_{g}$, with the forces acting also on symmetric systems while the torques are zero.

Before considering Model IV, we first consider two simpler models of Janus particles that assess the role
of various parameters appearing in the governing equations (Eqs.~\ref{eq.5} -~\ref{eq.9}). In
Section 4, results from Model 0 and Model I are presented. Model 0 is not truly a Janus particle: the cap is absent ($\delta=0$, homogeneous mass distribution) and the charge on the two faces is identical (i.e., $\psi_1=\psi_2$), making it impossible to distinguish them. Model I differentiates the role played by inhomogeneity in the surface charges ($\psi_1 \neq \psi_2$) of the particle and is the simplest representation of a Janus particle considered in this work. Note that both Model 0 and Model I represent particle behavior in the absence of shear flow, i.e., $\dot{\gamma}=0$ and therefore $V_x=0$.
In Section 5 the cap thickness is introduced, i.e., we consider $\delta>0$, while still keeping $\dot{\gamma}=0$ (hence $V_x=0$) and we refer to this formulation as Model II.
In Model II as well as in the subsequent models III and IV, where $\delta > 0$, the geometry of the cap is assumed to be thickest at the pole and becomes gradually thinner as it approaches the equator of the particle, which agrees well with recent experimental observations.\cite{rashidi2018local} The cap volume is estimated as the volume of a cylinder with a diameter of $2R$ of thickness $\delta$, $V_{\textsubscript{cap}}$ = $\pi$$R^2$$\delta$. Typically, Janus particles are coated with a thin layer of metal, e.g., platinum, which is denser than the SiO\textsubscript{2} core of the particle (e.g., $\rho_{Pt}=21.45$ vs $\rho_{SiO_2}=2.64 \,g/cm^{-3}$). Consequently, both the density mismatch and the geometry of the cap render the particle “bottom-heavy”.
Model III differs from Model II as the effect of shear flow  (i.e., $\dot{\gamma} \neq 0$) is included in the Langevin equation through a non-zero velocity term, $V_x$.
The ratio of the advection to Brownian forces is expressed by the Péclet number, $Pe_p = \dot{\gamma} R^2/D_0$, where $D_0$ is the translational diffusivity of the isolated Brownian particle in bulk. Depending on the particle size and strain rate, $Pe_p$ numbers vary over the range of $0 \leq Pe \leq 0.5 \times 10^4$ . Subsequently, the effect of the hydrodynamic friction is evaluated by inclusion of frictional parameters, $f_x$ and $f_{\theta}$, in the translational and rotational terms in Model IV.

Note that, in these equations the particle height ($z$) and position $(x)$, time ($t$), force ($F$), and the torque ($N$) are non-dimensionalized by  the Debye length ($\kappa^{-1}$),  the characteristic time ($T_{0}=\frac{3 \pi \mu R}{k_b T\kappa^2}$), thermal force ($k_b T/\kappa^{-1}$), and thermal energy ($k_b T$), respectively. Non-dimensionalized variables are indicated by a bar, e.g., $\bar{z}$.

\section{Methods}\label{sec:methods}

A BD simulation is conducted to evaluate the rotational and translational trajectories of a Janus particle by integrating the Langevin equations forward in time employing the explicit Euler time integration scheme. Thermal fluctuations are incorporated in the integration by sampling from the normal distribution $N(0,1)$. Note the standard white noise assumption for the Brownian force applies for the system studied here as the ratio of $\dot{\gamma}$ over the bath frequency $(\approx 10^{12}\,s^{-1})$ is less than $10^{-11}$. For larger ratios, i.e., very large $\dot{\gamma}$ or more viscous fluids, a non-white noise more accurately describes the random fluctuations~\cite{Pelargonio}. Each state point is integrated from time $t_0$ up to time $t_1$ and $N_p$ independent particles are simulated. Specific settings of numerical simulations are detailed in Section S2 of the ESI.\dag\,

Analysis of the trajectories results in PDFs that show the frequency with which the particle is found in a particular orientation and height at various conditions. In order to assess and minimize bias in the behavior of the Janus particle with respect to the initial positions and orientation, random initial position and orientations are utilized. Equilibrium behavior of the Janus particle is verified to be independent of the initial condition and the early-time stages of the dynamics are disregarded from the data analyzed, removing data from the beginning of the simulation up to the equilibration time $t_{eq}$ when necessary. See ESI Section S2\dag\, for details on the initial condition and on the equilibration time. This approach ensures that the results obtained studying different trajectories of the same Janus particle are statistically equivalent.

When the shear flow is included in the model, i.e., Model III, the focus is particularly on the rotational degree of freedom of the Janus particle. To characterize it, the average angular velocity of a each trajectory is measured by temporally averaging the velocity as
\begin{equation}
    \langle \omega \rangle_t = \frac{1}{t_1-t_{eq}} \int_{t_{eq}}^{t_1} v_{\theta}(\bar{t}) d\bar{t} \, ,
\end{equation}
where the instantaneous angular velocity $v_{\theta}(\bar{t})$ is simply estimated as $(\theta(\bar{t}+d\bar{t}) - \theta(\bar{t}))/d\bar{t}$.
The ensemble average of the angular velocity is then computed by averaging over different
simulations, obtaining a constant angular velocity value, $\omega$. The $\omega$ value is used as the order parameter of a dynamical transition between a rotating state, where $\omega>\omega_0$, and a non-rotating
state, where $\omega \leq \omega_0$.
The two states are distinguished by setting the threshold $\omega_0=0.05$ and verifying the robustness of the results by considering similar values of the threshold.

The trajectories, $\theta(\bar{t})$, exhibit a rich behavior, which is
classified using statistical testing.
Specifically, time is binned and an average value of the instantaneous velocity $v_{\theta}(\bar{t})$ is computed for each bin to obtain a time-dependent average angular velocity $\omega(t)$ (see ESI Section S2),\dag\, whose behavior is fitted to two functions: a constant and a sinusoidal.
A constant behavior of $\omega(\bar{t})$ indicates that the particle rotates with a velocity that is independent of its orientation and possibly does not rotate at all, if $\omega=0$; a sinusoidal behavior indicates that the rotation
is accelerated over time, an effect that is the consequence of the complex interplay between shear forces, gravitational attraction between the particle and the wall and electrostatic repulsion between the wall and the two sides of the Janus particle (which in general have different charge and hence experience different repulsion strength).
The best model among
the constant and the sinusoidal is selected using the Akaike Information Criterion,~\cite{burnham2002model} which properly accounts for the larger number of fitting parameters (four) that the sinusoidal model has over the constant (which has one). In this way, each trajectory is classified as being sinusoidal or constant, providing a number of classifications equal to the number of trajectories simulated per state point.
The more populated class is selected as the most representative of the state point.
Noting that the classification results in two possible classes for each state point, its reliability is further tested as follows.
The class is regarded as a binary variable and the null hypothesis that the outcome of the classification results from the sampling of a binomial distribution with
parameter $1/2$ and with $N_p$ trials is formulated.
The null hypothesis is tested setting the $p$ value to 0.01 and computing the probability of the classification
outcome under the null hypothesis.
The classification is regarded as reliable if such a probability is found smaller or equal to $p$. In this way, each state point is classified as sinusoidal, constant or is not classified at all (i.e., when the classification is regarded as not reliable).

\section{Dynamics of a Janus Particle Near a Wall}\label{sec:janus}

\begin{figure}[!htbp]
\centering
 \vspace{0 cm}
 \includegraphics[width=9cm]{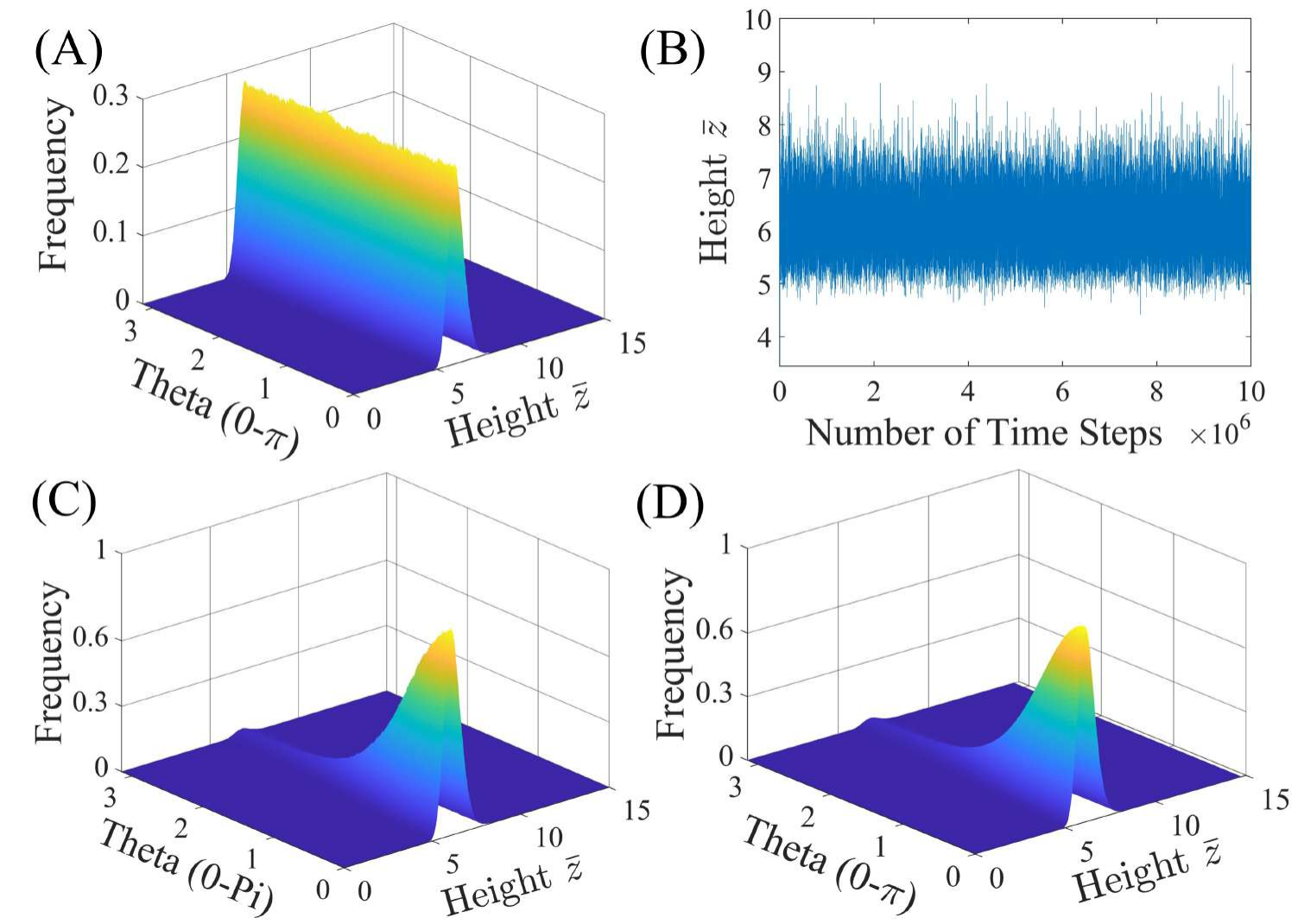}
 \caption{Impact of surface charge, $\psi$,  on the probability density function (PDF) for a Janus particle of $R=1 \, \mu m$ without a cap ($\delta =0$),  with surface charges, $\psi_1$ and $\psi_2$, interacting with a wall of surface charge, $\psi_w=-50 \,mV$, using Model 0: (A) PDF and (B) equilibrium height for $\psi_1=\psi_2=-20 \,mV$ and Model I: (C) PDF for $\psi_1=-20 \,mV$ and $\psi_2=-40\,mV$. (D) Boltzmann distribution prediction of the orientation distribution for particle simulated in (C).}
\label{fig2}
\end{figure}

The behavior of a homogeneously charged ($\psi_1 = \psi_2$) "Janus" particle of $R=1 \, \mu m$ without a cap ($\delta =0$) near a wall (SiO\textsubscript2) with negative surface charge polarity of $\psi_w = -50 \,mV$ is studied first to test the formulation and validity of the BD simulation, i.e., Model 0, and compared to a Janus particle without a cap ($\delta =0$), but with differing surface charges ($\psi_1 \neq \psi_2$), i.e., Model I, in Figure \ref{fig2}.

Figure \ref{fig2}A, shows the expected PDF with equal probability for all orientations for a "Janus" particle with uniform surface charge ($\psi_1=\psi_2=-20 \,mV$, Model 0) near a wall with $\psi_w=-50 \,mV$. The PDF is uniform because the driving force to rotate each face of the particle toward the wall is equal for both particle halves owing to the identical surface charge. The particle rotates around its center, $O$, at an equilibrium height of $\bar{z}$ = $6.19\pm0.45$ as shown in the height trajectory of the particle in Figure \ref{fig2}B.

The PDF shown in Figure \ref{fig2}C is obtained for a Janus particle where the surface potential of the two halves differ (Model I) and are specified as $\psi_1=-20 \,mV$ and $\psi_2=-40 \,mV$, while the wall surface charge is kept at $\psi_w=-50 \,mV$. Comparing the PDFs in Figures \ref{fig2}A and \ref{fig2}C, it is apparent that the particle stays at a similar equilibrium height but shows a bias in its configuration, i.e., the particle tends to orient the face of lower surface charge, face 1, toward the wall ($\theta =0$) owing to the greater potential gradient in surface charge. Calculation of the equilibrium height reveals $\bar{z} = 6.32\pm0.50$ slightly increased from the evenly charge particle. Careful inspection of the height distribution reveals that the particle is closer to the surface when face 2 is rotated away from the wall than when rotated toward the wall $(\bar{z}_{max,0} = 6.07$ vs $\bar{z}_{max,\pi} = 6.70)$ explaining the increased average equilibrium height value.

The dynamics of a Janus particle with the total potential interaction energy of $U_{tot}(z,\theta)$ should follow the Boltzmann distribution prediction for which the probability distribution of various configurations is given as $p_s(z,\theta) = A_p\cdot e^{(\frac{-U_{tot}(z,\theta)}{k_BT})}$ with $A_p$ chosen such that \(\int{p_s(z,\theta)}dzd\theta\)=1. Figure \ref{fig2}D shows the Boltzman distribution prediction for the same parameters used in the simulation of the PDF shown in Figure \ref{fig2}C. Comparison of Figures \ref{fig2}C and \ref{fig2}D shows agreement thereby inferring the accuracy of the BD simulation methodology used here. Note data shown in the ESI that this holds true also for the plain particle model (see ESI, Section S3).\dag\,

\section{Dynamics of a Bottom-Heavy Janus Particle Near a Wall}\label{sec:bh-janus}

After evaluating the accuracy of Models 0 and I for a uniformly charged "Janus" particle ($\psi_1 = \psi_2$) and the simplest Janus particle ($\psi_1 \neq \psi_2$), respectively, Model II introduces the effect of the cap, here made of platinum ($\rho_{Pt}=21.45 \,g/cm^3$), creating a bottom-heavy Janus particle. The weight of the cap and its geometry are incorporated in the simulations as described in Section 2. In this context, the cap of the particle (i.e., its bottom-heaviness) contributes to the rotational dynamics of the particle and tends to rotate the cap toward the wall to antiparallel align the particle director ($\pmb n_p$) with the wall's director ($\pmb n_w$) via the corresponding torque of $N_g$.

\begin{figure}[!htbp]
\centering
\includegraphics[width=9cm]{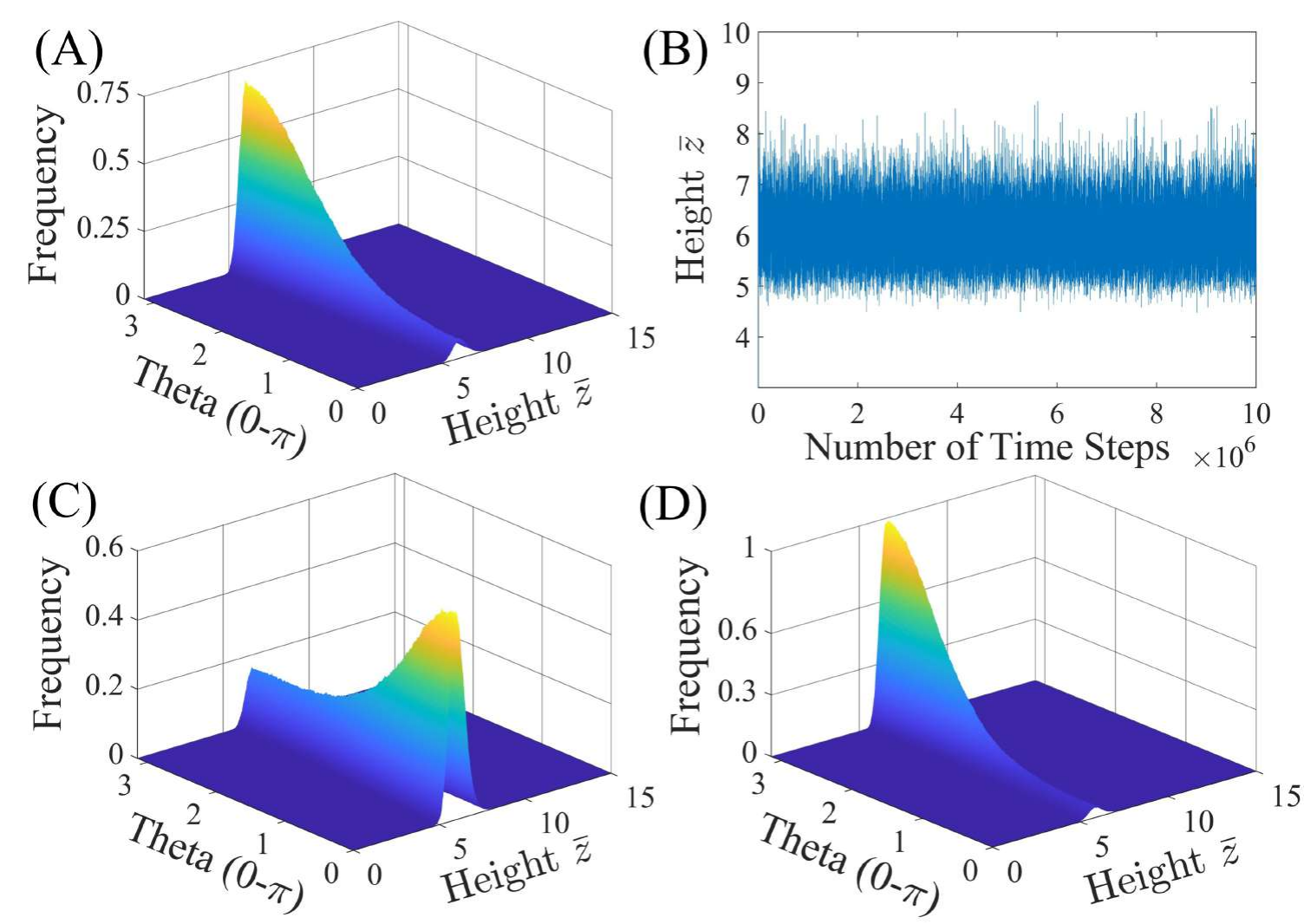}
 \caption{Impact of cap thickness and surface charge on the probability density function (PDF) of a bottom-heavy Pt-capped Janus particle of $R = 1 \,\mu m$ with cap thickness, $\delta$, and surface charges, $\psi_1$ and $\psi_2$, interacting with a wall of surface charge, $\psi_w=-50 \,mV$ using Model II: (A) PDF and (B) equilibrium height $\bar{z}$ for $\psi_1=\psi_2=-20 \,mV$ and $\delta=10\,nm$. (C) same as (A) with $\psi_2=-40 \,mV$. (D) Same as (C) with $\delta=30 \,nm$.\\}
\label{fig3}
\end{figure}

Figure \ref{fig3} shows the effect of incorporating a Pt cap ($ \delta > 0$) in the dynamical behavior of a particle with $\psi_1 = \psi_2$ and a Janus particle ($\psi_1 \neq \psi_2$) with varying $\delta$. Generally speaking, the bottom-heavy Janus particle stays at an equilibrium height and fluctuates with different preferred configurations near the wall owing to an interplay of the effects of the cap and surface charges.
The PDF in Figure \ref{fig3}A shows that for the case in which $\psi_1 = \psi_2$ (the wall potential is kept at $\psi_w=-50 \,mV$), the particle prefers rotating the face with the cap, face 2, toward the wall ($\theta=\pi$). Furthermore, according to the height trajectory (Fig. \ref{fig3}B), the bottom-heavy Janus particle fluctuates at an equilibrium height of $\bar{z} = 6.09\pm0.46$ slightly smaller (and therefore closer to the wall) compared to the uniformly charged "Janus" particle in Figure \ref{fig2}B  due to the additional gravitational force of the cap.

Figure \ref{fig3}C displays the PDF for the case in which the cap weight and the more negative surface potential on the same particle face, face 2, exert competing demands on the particle orientation; the cap forces face 2 towards the wall ($\theta=\pi$) due to the gravitational torque, while the more negative surface potential forces face 2 away from the wall ($\theta=0$). For $ \delta=10\,nm$, the surface potential gradient dominates over the cap torque, and the particle exhibits a preference for rotation of face 2 away from the wall. The average height is found to be $\bar{z} = 6.37\pm0.51$, with the preferred height for the cap-up configuration ($\theta=0$) peaking at $\bar{z} = 5.99$, while the particle in cap-down configuration ($\theta=\pi$)  is found most often at $\bar{z} = 6.70$, in good agreement with the stronger repulsion between the wall and face 2.
An increase in cap thickness to $\delta=30\,nm$ leads to the dominance of the gravitational cap torque that quenches the rotational motion of the particle to a mostly cap-down configuration ($\theta = \pi$) as shown in Figure \ref{fig3}D. The average equilibrium height of the particle during the simulation is $\bar{z} = 6.50\pm0.45$ with cap-up and cap-down configuration heights peaking at $\bar{z} = 5.80$ and $6.51$, respectively.
It is clear that the dynamic rotational behavior and the height of a bottom-heavy Janus particle depend strongly on the interplay of cap and surface properties. The complexity of this interaction is increased further by the introduction of shear flow and particle radius variations discussed in the next section revealing an intricate Janus particle flow behavior near a wall.

\section{Dynamics of a Bottom-Heavy Janus Particle Near a Wall Under Shear Flow}\label{sec:bh-janus-shear}

\begin{figure*}[!ht]
\centering
 \includegraphics[width=\textwidth]{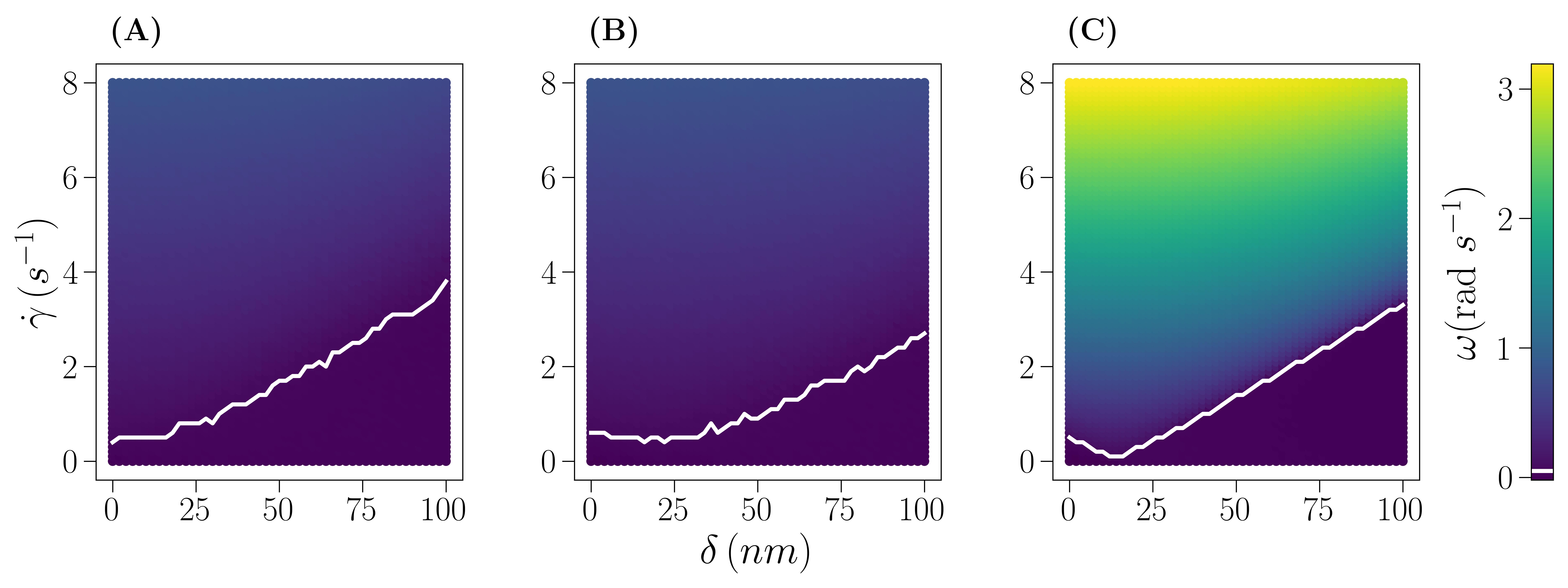}
 \caption{Angular velocity maps for a bottom-heavy Janus particle under varying shear flow ($\dot{\gamma}= 0-8 \,s^{-1}$) and cap thickness ($\delta = 0-100 \,nm$) using Model III for:
 (A) $R = 1 \mu m$, $\psi_1=-20 \,mV$,
 $\psi_2=-20 \,mV$ and $\psi_w=-50 \,mV$. (B) as in (A), but $\psi_2=-40 \,mV$.
(C) as in (B), but $R = 4 \mu m$. The color represents the average angular velocity as displayed by the color bar on right. The transition line (white) is drawn at $\omega=0.05 \,s^{-1}$.
 }
\label{fig4}
\end{figure*}

In Model III, the bottom-heavy Janus particle is subjected to shear flow with strain rate $\dot{\gamma}$ acting on the particle center, O.
Initial simulations are performed with a strain rate of $\dot{\gamma}= 6\,s^{-1}$ for sets of three Janus particles with $R=1\,\mu m$, $\delta = 86 \,nm$, $\psi_1 = -20 \,mV$, $\psi_2 = -40 \,mV$, and $\psi_w = -50 \,mV$. Data for translation in $\bar{x}$, $\theta$, and $\bar{z}$ as a function of $\bar{t}$ are provided for the three particles in the ESI Section S4 (Fig. S7, top left).\dag\,
The three traces shown for $\bar{x}(\bar{t})$ indicate that the Janus particles move forward in the direction of the flow.
The $\theta(\bar{t})$ traces show that each Janus particle starts in a unique, random orientation. Over time, $\theta$ values fluctuate between 0 and 2$\pi$ in an cyclic fashion implying that the Janus particle rotates from a cap-down orientation ($\theta = 0$) to a cap-up orientation ($\theta = \pi$) back to a cap-down orientation ($\theta = 2\pi$). Note that the particle orientation $\theta = 0$ is equivalent to $\theta = 2\pi$ resulting in the abrupt change in $\theta$ at the end of each cycle. The $\bar{z}(\bar{t})$ traces closely follow the $\theta(\bar{t})$ traces with the particles at a high $\bar{z}$ when the particles are cap down and a lower $\bar{z}$ when the particles are cap up.
A change in particle radius to $R=4 \,\mu m$ at the same $\dot{\gamma}= 6\,s^{-1}$ and cap thickness, and for the same simulation length shows the expected longer distance $\bar{x}(\bar{t})$ and overall smaller $\bar{z}$ values with the oscillations in $\bar{z}$ more clearly matching the particle rotation (ESI,\dag\, Fig. S7, bottom left).
Decrease of $\dot{\gamma}$ to $1s^{-1}$ (ESI,\dag\, Fig. S7, top and bottom right) leads to a different $\theta$ behavior for both Janus particle sizes, where the Janus particle rotates until it assumes a constant $\theta$ value, i.e., the strain rate is not strong enough to induce
rotation ($\omega < \omega_0$) and the particle slides along the wall with constant $\theta$.
The distinct behavior observed in the cap rotation, i.e., $\theta(t)$, between the two particle sizes and as a function of $\dot{\gamma}$ prompts a careful search of the $\dot{\gamma}$,$\delta$-parameter space for $R=0.75, 1, 2,$ and $4\,\mu m$ with $\psi_1{/}\psi_2{/}\psi_w$ combinations of $-20{/}-20{/}-20\,mV$, $-20{/}-20{/}-50\,mV$, $-40{/}-20{/}-20\,mV$, $-40{/}-20{/}-20\,mV$, $-20{/}-40{/}-20\,mV$, and $-20{/}-40{/}-50m\,V$ (see ESI Section 5, Figs. S8-S12). The findings from this parameter scan are summarized here.

Figure~\ref{fig4} displays the angular velocity $\omega$ (see Section \ref{sec:methods}) of a Janus particle with the same $\psi_1$, $\psi_2$ and $\psi_w$ values considered in Section 5 (Fig.~\ref{fig3}) for the same particle radius $R=1 \mu m$ (Fig.~\ref{fig4}A and B) and for a larger particle with radius $R=4 \mu m$ and surface charges $\psi_1=-20 \,mV$, $\psi_2=-40 \,mV$ and $\psi_w = -50 \,mV$ (Fig.~\ref{fig4}C) for cap thicknesses $\delta = 0-100\,nm$ and strain rates $\dot{\gamma}= 0-8 \,s^{-1}$.
All systems display the transition between a rotating ($\omega > \omega_0$)  and a non-rotating behavior ($\omega < \omega_0$) described above as marked by the white line in Figure~\ref{fig4}. Above the transition line, $\omega$ increases with increasing $\dot\gamma$ at constant cap thickness $\delta$, whereas at constant $\dot\gamma$ and increasing $\delta$, the change in $\omega$ is non-trivial.
As the distance between the particle center, O, and the top of the cap, i.e., the lever arm, grows linearly with $R$ at given $\delta$, the torque acts more effectively on
large particles, leading to the higher values of $\omega$ observed at the larger $R$ (see colors in Fig.~\ref{fig4}A/B vs C).

The onset of a non-negligible angular velocity at $\delta=0$ follows the expected trends by increasing from Figure~\ref{fig4}A to B due to the higher surface charge asymmetry and decreasing from Figure~\ref{fig4}B to C owing to the increase in particle radius. However, once the cap is introduced ($\delta > 0$) and increased in thickness, the transition line shows a distinct behavior for each system, ranging from monotonic increase (Fig.~\ref{fig4}A) to slight decrease followed by a slow increase (Fig.~\ref{fig4}B) and sharp decrease followed by a linear increase.
\begin{figure*}[!ht]
\centering
 \includegraphics[width=\textwidth]{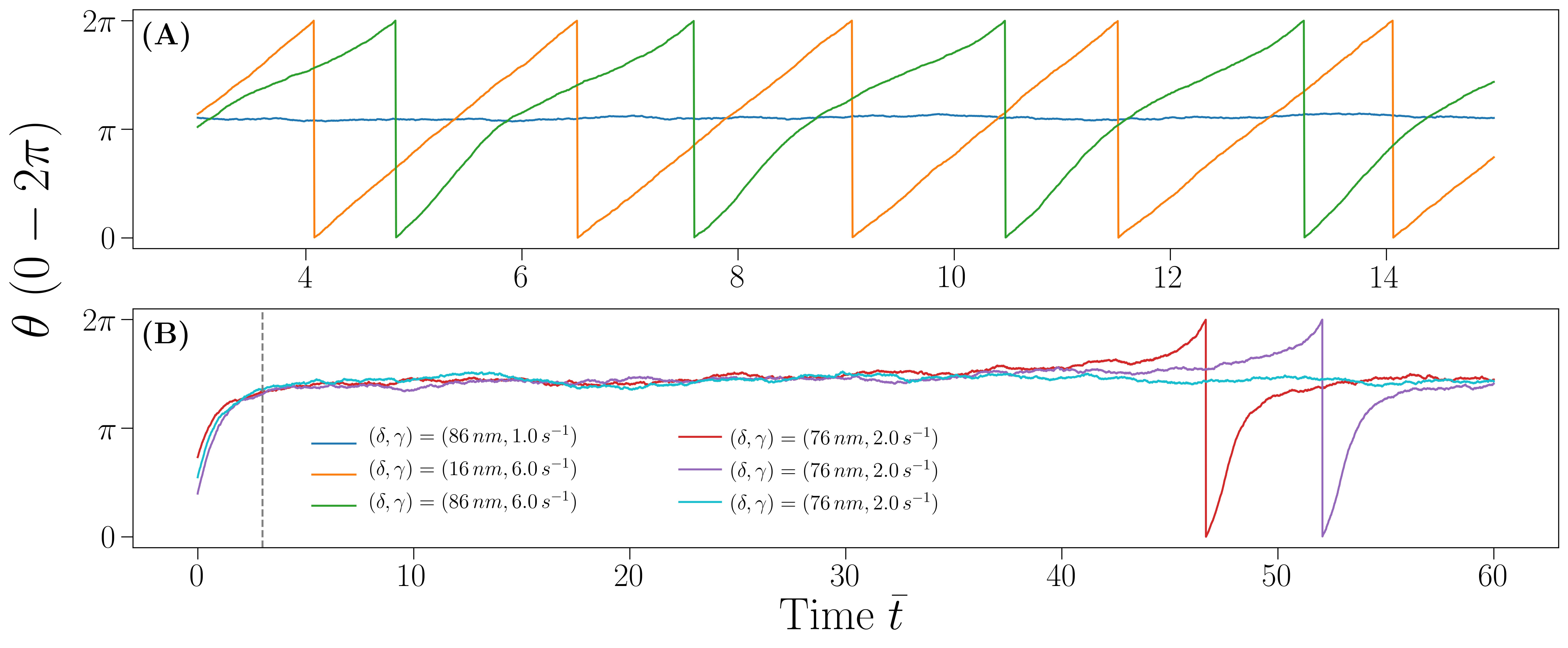}
 \caption{Representative angular trajectories for a system with
 $R=4 \, \mu m$, $\psi_1 = -20 \, mV$, $\psi_2 = -40 \, mV$, and $\psi_w = -50 \, mV$.
 (A) Angular trajectories in a short-time window for state points specified in the legend shown in (B).
 (B) Angular trajectories in a long-time window for systems whose parameters are given in the legend.
 Note that the three trajectories shown in (B) correspond to different simulations of the same system.
 The dashed vertical line is the equilibration time after which the data are analyzed.
 }
\label{fig5}
\end{figure*}
The transition line, a locus of points in the diagram of the form $(\delta, \dot{\gamma})$, fully characterizes the transition: above (below) the locus the system is in the rotating (non-rotating) state.
For example, if two points of the locus $(\delta_1, \dot{\gamma}_1)$ and $(\delta_2, \dot{\gamma}_2)$ are chosen such that the system is at the transition at those points, it is true for (almost) all systems that if
$\delta_1 < \delta_2$, then $\dot{\gamma}_1 < \dot{\gamma}_2$.
In other words, it is (almost) always true that increasing the cap thickness requires a stronger shear flow for the rotating phase to be entered (see Fig.~\ref{fig4}A and ESI Section S5).\dag\,
The only exception to this behavior is observed for Janus particles whose heavy side (face 2) is more repelled ($\psi_2 = -40\,mV$) by the wall then the light side (face 1, $\psi_1 = -20\,mV$) as shown in Figure~\ref{fig4}B/C (see also ESI Section S5, Figs. S8 and S9).\dag\,
In this case, it is observed that $\delta_1 < \delta_2$ results in $\dot{\gamma}_1 \geq \dot{\gamma}_2$ for sufficiently small $\delta$ (lower left corner).
This behavior is persistent across different particle sizes and is more discernible for particles with $R \geq 2$. A different way of describing this phenomenon is
to see the locus of points as a set of values of $\dot{\gamma}$ that is a function of $\delta$.
In this perspective, it can be stated that $\dot{\gamma}$ is (almost) always a growing function of $\delta$, with an exception at small $\delta$ when the charge asymmetry ($\psi_1$ less negative than $\psi_2$) makes the heavy face, face 2, more repelled by the wall.
A minimum in the transition line then represents the cap thickness at which the opposing cap torque and surface charge gradient balance each other.

Surprisingly, the location of the non-rotating region depends only weakly on the particle radius, charge balance, and cap thickness.
This behavior is explained by taking into account the complex interplay between particle radius, surface potential, and cap thickness, which together determine - for a given strain rate - the average distance $\bar{z}$ between the particle and the wall (see ESI Section S5,\dag\, Fig. S10).
For example, the strain rate is less effective when a more negative $\psi_2$ than $\psi_1$ results in a low $\bar{z}$,  but is more effective when a large R and/or a strong asymmetry in the particle's mass distribution due to a large $\delta$ result also in a low $\bar{z}$.

At fixed $R$ (Fig.~\ref{fig4}A vs B), an increase in electrostatic repulsion only slightly shifts the transition line to lower $\dot\gamma$ values, as it pushes the bottom-heavy particles farther from the wall, thus favoring their rotation. On increasing $R$ (Fig.~\ref{fig4}B vs C), the lever arm grows, favoring the rotating behavior at a given $\dot \gamma$, but such an increase is compensated by an increase in particle weight, which brings the particle closer to the wall, thus disfavoring the particle rotation due to the electrostatic repulsion. The previous argument helps to clarify the aforementioned complex dependence of $\omega$ on $\delta$ at constant $\dot \gamma$. In particular, a non-monotonic behavior is observed when the repulsion strength between the wall and the heavy side of the particle is strong. This is because, as a consequence of the electrostatic repulsion, the particle is further away from the wall when the heavy side is closer to it. This effect interplays with the previously noted ones and triggers the rotation of particles with thin  caps (including the $\delta=0$ case) at values of $\dot \gamma$ that would not be sufficient to trigger it in absence of the charge asymmetry: the charge asymmetry pushes the particle upward enough to allow $\dot \gamma$ to act more effectively on it, hence triggering the rotational behavior.
\begin{figure*}[!ht]
\centering
 \includegraphics[width=\textwidth]{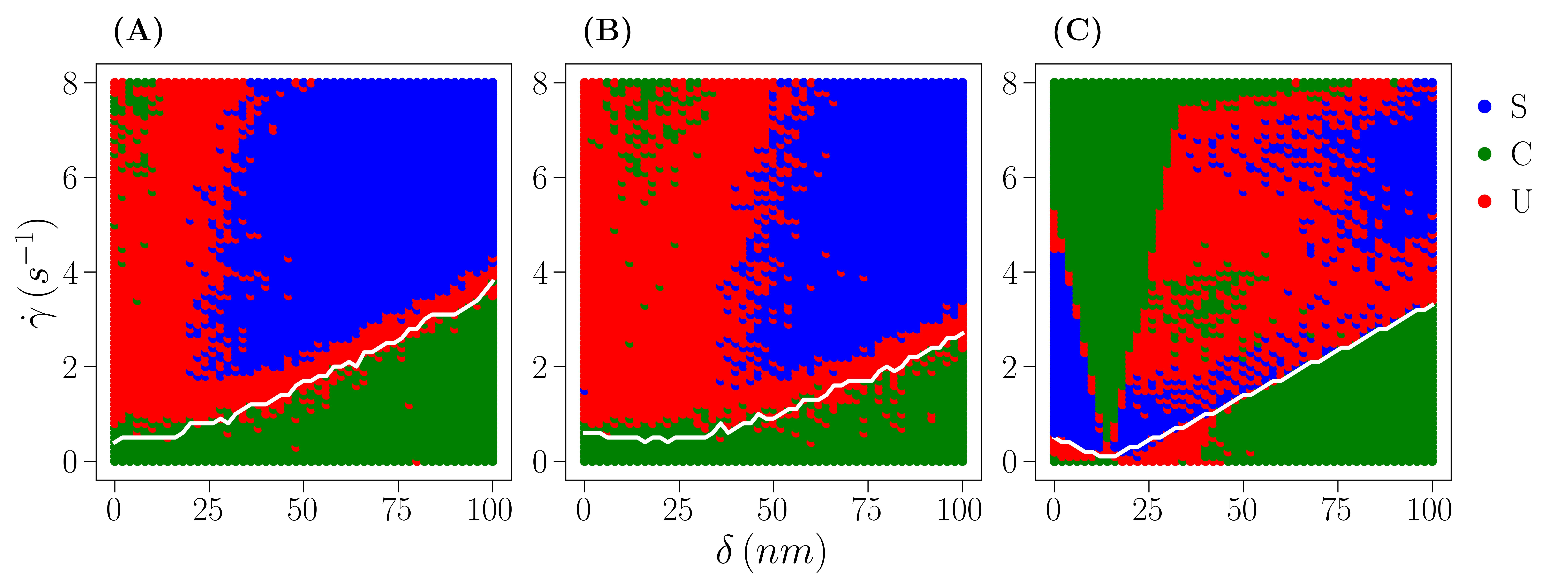}
  \caption{State diagram of a bottom-heavy Janus particle under varying shear flow ($\dot{\gamma}= 0-8 s^{-1}$) and cap thickness ($\delta = 0-100\,nm$) for: (A) $R = 1 \,\mu m$, $\psi_1=-20 \,mV$,
 $\psi_2=-20 \,mV$ and $\psi_w=-50 \,mV$. (B) Same as in (A), but $\psi_2=-40 \,mV$.
(C) Same as in (B), but $R = 4 \,\mu m$
The white line is the same transition line drawn in Figure~\ref{fig4}, colors represent
 the classification of each state point as described by the legend on the right. Symbols in the legend stand for Sinusoidal (S), i.e., non-consntant angular velocity, Constant (C), i.e., constant angular velocity, and Unclassified (U).
 }
\label{fig6}
\end{figure*}

The phenomenology is, however, richer than what can be deduced from Figure~\ref{fig4} only.
A visual inspection of individual trajectories (Fig.~\ref{fig5}) for the same system under different conditions shows evident differences.
All curves are for $R=4 \, \mu m$, $\psi_1 = -20 \, mV$, $\psi_2 = -40 \, mV$ and $\psi_w = -50 \, mV$, i.e., they correspond to state points in the angular velocity map shown in Figure~\ref{fig4}C.
In Figure~\ref{fig5}A, the curve for $\delta  = 86 \, nm$ and $\dot\gamma = 1.0 \, s^{-1}$ (blue) is an example of a trajectory sampled in the non-rotating phase and indeed the particle orientation simply experiences Brownian fluctuations.
On increasing the shear rate to 6.0 $s^{-1}$ while reducing the cap thickness to 16 $nm$ (orange), the particle enters the rotating phase.
The trajectory shows a growth of $\theta$ with time that is well approximated by a linear function, indicating a constant angular velocity.
On increasing the cap thickness back to 86 $nm$ (green), the behavior of $\theta(\bar{t})$ changes significantly: the first half of the rotation, for $0 \leq \theta < \pi$ is much faster than the second half.
This change in angular velocity happens because the cap is heavy so gravity tends to keep it close to the wall and more work is required for the torque to fully rotate a particle in the opposite orientation, where the heavy side is aligned with the wall normal $\pmb n_w$ ($\theta = 0$).
This behavior is ascribed to an oscillating trend of the angular velocity, which is higher in the first part of the rotation and lower in the second.
The observation motivates the classification introduced in Section \ref{sec:methods}.
Figure~\ref{fig5}B illustrates a difficulty encountered by
the classification. The curves in Figure~\ref{fig5}B show a mixed behavior despite corresponding to different simulations of the same state point in the angular velocity map.
One of the curves (red) shows a single rotation over the whole temporal window of the numerical integration ($\bar{t}\approx 47$), the purple curve also shows a single rotation, but almost at the end of the integration ($\bar{t}\approx 52$) and the cyan curve does not show a rotation at all, despite the fact that a rotation would be observed if a longer numerical integration time were used.
However, there is not a length of the numerical integration time that guarantees observation of at least one rotation for systems that exhibit rotations and not to observe rotation at all for systems that do not undergo rotations.
When the model parameters are such that the rotational behavior exists, but the angular velocity is small, two different scenarios can be observed: very slow rotations with constant angular velocities, or the limit behavior of the sinusoidal angular velocity previously described.

Figure~\ref{fig5}B is an example of such a limit behavior: as shown by the red curve, the time it takes to complete the first half of the rotation is extremely short, due the heavy cap driving face 2 toward the wall, while the time it takes to complete the second half of the rotation is exceedingly large as shear flow and charge repulsion must overcome the gravitational torque from the cap.
This phenomenon can be described by the first passage time distribution as a parametric function of
the orientation $\theta$:~\cite{gardiner2009handbook} such a function describes the probability that $\theta$ reaches a certain (parametric) value in a time $t^*$, which is the random variable of the distribution.
Stochastic processes near a phase transition are typically characterized by a fat tailed distribution of the first passage time~\cite{DiSanto2017Simple} and the results presented here suggest that the system under analysis makes no exception: for a large value of the parameter $\theta$, the distribution would likely appear heavy tailed, indicating that no matter how long the numerical integration is, the probability to not observe a full rotation would still be non-zero.

Note the regime of constant angular velocity can also be seen as a purely rotational regime, where the translational motion of the particle is strongly correlated with it's orientation, as one expects for a ball rotating on a plane, while the variable angular velocity, induced by charge imbalance and asymmetry of the mass distribution, can be seen as a type of motion in which the correlation between orientation and position along the plane is increasingly weaker as the linearity in $\theta(t)$ is lost, due to a ''slipping" behavior in which the particles translate while rotating independently of the translating motion.

The results of the classification are shown as state diagrams in Figure~\ref{eq.6} and in the ESI Section S6.\dag\, The transition line between the rotating and the non-rotating behavior is identified by the white transition line.
Almost all the state points below it are ascribed to constant angular velocity (green state point) with a small region of unclassified states (red state point) inside the non-rotating phase only for large particle radius, small cap thickness, and small shear flow.
Just above the transition line, a region of unclassified states points is systematically observed. This is due to the phenomenon exemplified in Figure~\ref{fig5}B and explained in the previous paragraph: near the transition, some realizations appear as constant (cyan curve) and some are clearly not well fitted by a constant function (red and purple curves), resulting in state points that are not classified at all.
Further above the transition line, at large cap thickness and strong shear flow, the system is systematically characterized by a sinusoidal angular velocity (blue state point).
On progressively reducing the cap thickness, the system enters first a region of unclassified state points in which the distinction between the two behaviors is not simple as the difference in slopes between the two regimes of rotation is extremely small, but not small enough for the state points to be systematically ascribed to a constant behavior. On further reducing the cap thickness, a region of constant angular velocity is entered.
This region displays an interesting behavior as the model parameters change: on reducing $R$, it systematically shifts to the left (Fig. \ref{fig6}C vs B and ESI, Fig. S12),\dag\, it is observed to shift to the right (Fig. \ref{fig6}A vs B and ESI, Figs. S11 and S12)\dag\, if the charge of the heavy side of the particle is increased (in absolute value, at fixed $R$) and it shifts upward if the other side is overcharged with respect to the wall (ESI, Fig. S11).\dag\, See ESI Section S6\dag\, for a more detailed analysis of the results.

Despite being apparently intricate, this behavior is relatively simple to understand.
Constant but sinusoidal angular velocity is observed when the shear rate is strong enough to induce rotational motion (above the transition line), but the charge and density imbalances are not so strong that one side is more prone to be closer to the wall than the other.
Hence, at comparable repulsion between the two sides and the wall ($\psi_1 = \psi_2$), only the cap induces asymmetry and this can be compensated when the cap is small and the shear flow is large.
Reducing $\psi_2$ to more negative potentials means that the light side of the particle is more likely to stay close to the wall, which is balanced by the cap pushing the particle in the opposite orientation, hence increasing the width of the constant regime.
When the light side is strongly repelled by the wall, the region shifts upward as a given asymmetry in the mass distribution requires a strong shear rate to compensate the repulsion at small cap sizes.
Further, when the constant behavior region is observed at large cap thickness, it is also seen that a new sinusoidal region exists on its left, when both cap thickness and shear rate are small (ESI, Figs. S11 and S12, bottom row)\dag\,.

Summarizing the behavior of bottom-heavy Janus particle under shear flow, it can be stated that outside of the rotating region the system is characterized by a sinusoidal angular velocity if the cap is thick enough to induce a strong asymmetry. Reducing the cap thickness at a given shear rate implies reducing the variability in $\omega(\bar{t})$ up to completely balancing the speed of rotation in the two half periods and, possibly, on further reducing the cap thickness a sinusoidal angular velocity is observed again (especially for large particles), as the charge imbalance is not compensated by the weak flow. Data provided in the ESI (Section S5 \& S6)\dag\, shows that this behavior is persistent across different model settings.

Last but not least, in Model IV, the impact of hydrodynamic friction is considered taking into account the Janus particle's closeness to the wall for the $R=1 \, \mu m$ and $R=4 \, \mu m$ systems discussed at the beginning of this section ($\delta = 86 \,nm$, $\psi_1 = -20 \,mV$, $\psi_2 = -40 \,mV$, and $\psi_w = -50 \,mV$). Based on Goldman et al.’s report,\cite{goldman1967slow} frictional factors $f_x$ and $f_{\theta}$, which are functions of $\bar{z}$, are determined and introduced into Eqs.~\ref{eq.5} and ~\ref{eq.7}, respectively. For a bottom-heavy Janus particle of $R = 1\,\mu m$ and cap thickness of 86 nm at an average height of $\bar{z} = 6 \pm 1$, $f_x=0.9887$ and $f_{\theta}=0.9882$ indicating that these particles experience very little hydrodynamic friction. The situation changes when the particle size is increased to $R = 4\,\mu m$ and the particles display a average $\bar{z} = 3 \pm 0.5$ resulting in an  $f_x = 0.8342$ and $f_{\theta} = 0.8428$. Inclusion of these friction factors in the simulation reduces the angular velocity of the $R = 4\,\mu m$ Janus particle resulting in approximately 20\% fewer rotations as shown in the ESI Section S7, Figure S13,\dag\, while the $R = 1\,\mu m$ Janus particle is too far away from the surface to be substantially impacted by friction.

\section{Conclusion}
The behavior of a bottom-heavy, charged Janus particle near a charged wall under shear flow is fully rationalized, providing insights into how both confinement and shear flow can be utilized to manipulate the particle's translational and rotational motion.

The dynamical behavior of a Janus particle is described using a model based on gravitational and electric double layer interactions. The simplest Janus particle model, Model I, shows that the Janus particle without a cap of differing density tends to orient toward the wall with the face of lower surface potential because of the greater potential gradient. For the case of a uniformly charged bottom-heavy Janus particle, i.e., a particle with a cap of mismatched density but uniform surface charge, Model II, the cap quenches the rotational motion of the particle leading to a preference for a cap-down orientation adjacent to the wall in good agreement with existing literature.\cite{doi:10.1063/1.4994843, PhysRevE.101.042606} Adjustment of the cap’s surface charge leads to a competition between gravitational torque and electrostatic repulsion in which the position of the particle above the wall and its rotational behavior become a function of radius, surface potential, and cap thickness.

The relationship is further complicated by the introduction of shear flow at small $Pe_p$ in Model III. Thorough analysis of the particle behavior for a range of shear flows ($0-8 \,s^{-1}$), cap thicknesses ($0-100 \,nm$), radii ($0.75-4 \,\mu m$), and surface potential combinations ($\psi_w = -20, -50 \,mV$ and $\psi_{1,2} = -20, -40 \,mV$) reveals complex angular velocity profiles for the bottom-heavy Janus particle with regions of no rotation, constant rotation, and time-varying rotation.
Use of statistical analysis leads to state diagrams that reveal $\delta, \dot \gamma\,$-combinations for which the passive, bottom-heavy Janus particle shows constant angular velocity with and without rotation and a sinusoidal angular velocity. The latter state can be described as the particle reaching a cap down orientation followed by a sliding event before completing its rotation where the sliding time can be engineered and therefore is of potential interest when patch-wall interaction needs to be timed in applications such as drug delivery or environmental remediation.
Finally, we note that hydrodynamic friction, Model IV, effectively reduces the particle's ability to rotate by $\approx20\%$ when the particle is close to the wall, which occurs for large radii and/or thick caps at high shear rates, but can be neglected for small particles and/or thin caps.~\nocite{Volpe:09}

\section*{Author contributions}
Z.J., U.C., and I.K. discussed and conceptualized the project. Z.J. formulated the theoretical model, developed the initial code and provided a first draft of the manuscript as part of her Ph.D. Thesis.\cite{Jalilvand} I.K. revised the theoretical model and finalized the code. Numerical simulations were performed by Z.J. I.K. and D.N.. D.N. conceptualized, implemented, and performed the statistical analysis needed for the angular velocity maps and state diagrams. D.N., E.B., and I.K. discussed the interpretation of angular velocity maps and state diagrams and revised the manuscript during I.K.'s sabbatical stay in E.B.'s group. All authors have read and approved the final version.

\section*{Data availability}
Data supporting the statements and conclusions made in the manuscript are included as part of the ESI.\dag\,
Additional data is available upon reasonable request from the corresponding authors.

\section*{Conflicts of interest}
There are no conflicts to declare.

\section*{Acknowledgments}
This work was supported by the National Science Foundation under award CBET-1705565. I.K. and U.C. also acknowledge support from NSF grant number ESE-2112550. I.K. thanks Eric McPherson and Justin Pinca for helpful discussions. D.N. and E.B. acknowledge support from the Austrian Science Fund (FWF) under Project. No. Y-1163-N27.



\section{Supplemental Information}

\renewcommand{\figurename}{SFigure}
\renewcommand{\thefigure}{S\arabic{figure}}
\renewcommand{\thesection}{S\arabic{section}}
\renewcommand{\thetable}{S\arabic{table}}
\renewcommand{\theequation}{S\arabic{equation}}

\setcounter{figure}{0}
\setcounter{table}{0}
\setcounter{equation}{0}
\setcounter{section}{0}

The supporting information document provides additional details on the model for a plain, charged particle near a charged wall (Section \ref{Sec:S1}), simulation details for data presented in the manuscript (Section \ref{Sec:S2}), validation of the model introduced in Section \ref{Sec:S1} (Section \ref{Sec:S3}), impact of shear flow on particle behavior (Section \ref{Sec:S4}), angular velocity maps (Section \ref{Sec:S5}), state diagrams (Section \ref{Sec:S6}), and data illustrating the impact of hydrodynamics (Section \ref{Sec:S7}).

\section {BD Simulation Details for a Single Charged Particle Near a Charged Wall} \label{Sec:S1}

A spherical particle of radius $R$ is considered.
The particle has a uniform surface charge $\psi_p$, is suspended in a density-mismatched Newtonian fluid with viscosity $\mu$, an electrostatic permittivity $\varepsilon_0$, and an inverse Debye screening length $\kappa$ that is determined by the ionic strength $I$ of the fluid.
The thermal energy of the system is represented by $k_B T$, where $k_B$ is Boltzmann’s constant.
Importantly, the  particle is bounded by a charged wall ($\psi_w$).
Assuming a constant charge condition and using height $z$ to describe the separation distance of  the particle surface from the wall, the surface interaction energy per unit area, $W$, resulting from the electric double layer interaction, is expressed by equation~(\ref{eqn.1}):\cite{doi:10.1021/acs.langmuir.5b03611}
\begin{equation}
     W=\varepsilon \kappa \frac{2\psi_p\psi_w e^{-\kappa z} +(\psi^2_p+\psi^2_w) e^{-2\kappa z}}{1-e^{-2\kappa z}} \,.
 \label{eqn.1}
\end{equation}
Equation~(\ref{eqn.1}) is based on the Poisson-Boltzmann theory.
The force, $F$, between two interacting objects, can be related to the surface interaction energy, $W$, via the Derjaguin approximation, equation~(\ref{eqn.2}):
\begin{equation}
F=2\pi R_{eff} W \,,
\label{eqn.2}
\end{equation}
where the effective radius, $R_{eff}$, for the case of a sphere interacting with a planar substrate is equal to the particle radius $R$. Integrating equation~(\ref{eqn.2}) over the height $z$, provides the electrostatic double-layer interaction potential, $U_{EDL}$, as shown in equation~(\ref{eqn.3}).

\begin{equation}
\begin{split}
     U_{EDL} & =2\pi R \varepsilon \bigg[ \psi_p\psi_w \ln{\frac{1+e^{-\kappa z}}{1-e^{-\kappa z}}}- \\
     & (\psi^2_p+\psi^2_w)\frac{\ln({1-e^{-2\kappa z}})}{2} \bigg]
\end{split}
 \label{eqn.3}
\end{equation}

A theoretical model is employed on the basis of the stochastic Langevin equation to probe the dynamics of the particle interacting with the wall according to equation~(\ref{eqn.4}).

\begin{equation}
     \frac{dr_i}{dt} = u^{\infty}(r_i) + \frac{1}{\gamma} (F^B_i + F^{nh}_i)
 \label{eqn.4}
\end{equation}
Here, $u^{\infty}(r_i)$ is the velocity of the ambient fluid at particle position \{$r_i$\} and $\gamma$ is the drag coefficient for a perfect sphere, $6\pi \mu R$. The non-hydrodynamic force, $F^{nh}_i$, includes any surface forces or external body forces, i.e., inter-particle interaction forces, that are a function of \{$r_i$\}. $F^B_i$ is a random fluctuating force acting on the particle that describes Brownian diffusion and renders equation~(\ref{eqn.4}) a stochastic differential equation. $F_i^B$ is assumed to represent a Gaussian stochastic process with the following moments:

\begin{equation}
     \{F^B_i\} = 0 \,,
 \label{eqn.5}
\end{equation}
\begin{equation}
     \langle\{F^B_i(t)\}\{F^B_i(t')\}\rangle = 2\gamma k_BT\delta(t-t') \,,
 \label{eqn.6}
\end{equation}
\\
where $T$ is the absolute temperature and $\delta(t-t')$ is the Dirac delta function. The magnitude of $F^B_i$ is a function of $T$ and the drag force, which follows the fluctuation-dissipation theorem. For a colloidal particle, $F_i^B$ mimics the impact of the ambient fluid’s fluctuating molecules on the particle and is expressed by the white noise function $\eta(t)$, which is a Gaussian stochastic process with moments $\langle\eta(t)\rangle$ = 0 and $\langle\eta(t)\eta(t')\rangle$ = $\delta(t-t')$. 
Note the standard white noise assumption for the Brownian force applies for the system studied here as the ratio of $\dot{\gamma}$ over the bath frequency $(\approx 10^{12}\,s^{-1})$ is less than $10^{-11}$. For larger ratios, i.e., very large $\dot{\gamma}$ or more viscous fluids, a non-white noise more accurately describes the random fluctuations~\cite{Pelargonio}. Additionally, the ratio of the advection to Brownian forces is expressed by the Péclet number based on the colloidal Brownian/diffusion relaxation, $Pe_p= \dot{\gamma} R^2/D_0$, where $D_0$ is the translational diffusivity of the isolated Brownian particle in bulk. Depending on the particle size and strain rate, $Pe_p$ numbers vary over the range of $0 \leqslant Pe_p \leqslant 0.5 \times 10^{-4}$.
The sedimentation force and particle-wall interactions are incorporated into the model via equation~(\ref{eqn.4}), within the framework of the Derjaguin-Landau-Verwey-Overbeek (DLVO) theory as non-hydrodynamic forces acting on the particle, $F^{nh}$, to describe the dynamical behavior of the particle.

\section {Details of numerical simulations and of statistical testing} \label{Sec:S2}
Results presented in Figures 2 and 3 in the manuscript are obtained from Matlab simulations of $N_p = 30$ particles over $\bar t =1\times 10^4$  and $dt = 0.001$ resulting in $1 \times 10^7$ integration steps. Owing to the large number of integration steps and the short equilibration time (ca. $500$ steps) needed for the particle, all orientations and heights are included in the averaging.  The initial condition is randomized using Matlab's default seed by assigning to $z$ a random value from a normal distribution with mean 3 and variance 1, while the initial condition on the particle orientation $\theta$ is drawn uniformly in the interval $[0, \pi]$. $z$ and $\theta$ axes are divided into 800 and 600 bins, respectively.

Results presented in Section 6 of the manuscript are obtained by simulating $N_p=50$
independent particles, each integrated from $t_0=0$ to $t_1=60$ and using $dt=0.006$, resulting
in $1\times 10^4$ integration steps.
The equilibration time is set to $t_{eq}=3$, which corresponds to removal of the first 500 integration steps from the data analyzed.
The initial condition is randomized by assigning to $x$ and $y$ a random value from a normal distribution with mean 3 and variance 1, the initial condition on the particle orientation $\theta$ is drawn uniformly in the interval $[0, \pi]$.
Note, however, that the system quickly reaches equilibrium so the dynamics observed at $\bar{t} \geq t_{eq}$ is completely independent from the initial condition.

Angular velocity maps (see Section \ref{Sec:S5}) are obtained by computing the integral in equation~(10) of the manuscript and averaging over the 50 particles simulated.
For state diagrams (see Section \ref{Sec:S6}), the period of the rotation of each particle is computed as $T=2\pi/\langle \omega \rangle_t$. Subsequently, the number of
bins is set to $\floor{12 (t_1-t_{eq}) / T}$, corresponding to 12 bins per period,
along the entire integration time.
For particles with very small $\langle \omega \rangle_t$, i.e., very large periods, this results in a very small number of bins.
If the number of bins is smaller than 100, it is manually set to 100.
Note that the results of the classification might slightly depend on the
assignment of the number of binds, especially in regions where the two
behaviors are close.
The resulting trajectories of the orientation averaged over time are used to compute the derivative to obtain the time-dependent angular velocity $\omega(\bar{t})$.

\begin{figure*}[!ht]
\centering
\includegraphics[width=\textwidth]{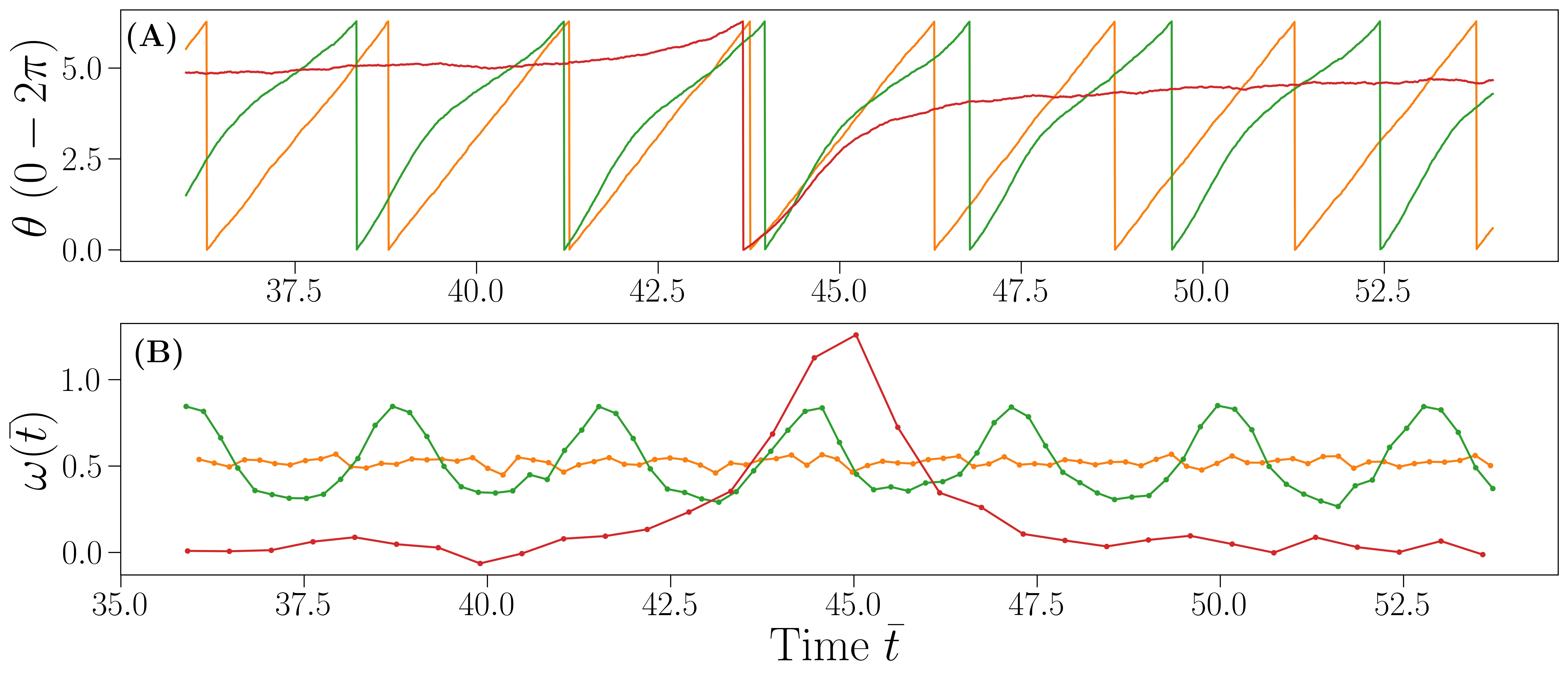}
 \caption{Representation of angular trajectories and corresponding time-dependent
 angular velocity.
 Top: angular trajectories for three states (orange and green from Figure~5A and red from Figure~5B) shown in Figure~5 of the manuscript.
 All curves are for $R=4 \, \mu m$, the orange curve is for $\delta=16 \, nm$ and
 $\dot{\gamma} = 6.0 \, s^{-1}$, the green curve is for $\delta=16 \, nm$ and
 $\dot{\gamma} = 2.0 \, s^{-1}$ and the red curve is for $\delta=76 \, nm$ and
  $\dot{\gamma} = 2.0 \, s^{-1}$.
  Bottom: Time dependent angular velocity obtained from the three trajectories in top panel using the same color scheme.}
\label{fig:fit_traject}
\end{figure*}

\begin{figure*}[!ht]
\centering
 \includegraphics[width=\textwidth]{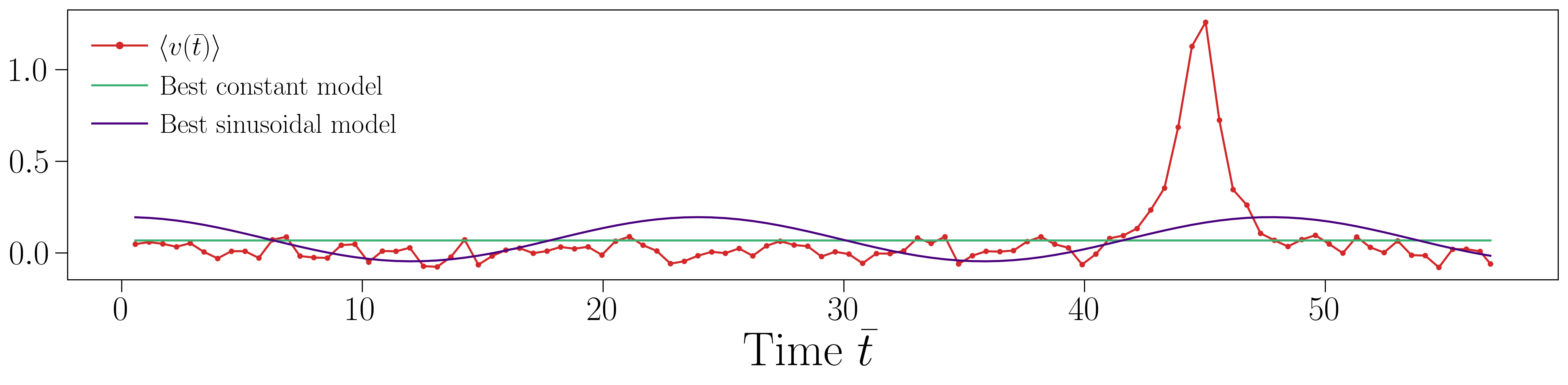}
 \caption{Fit to a trajectory at the boundary between rotating and non-rotating
 behavior. The red curve is the same data as the red curve in Figure~\ref{fig:fit_traject}, i.e., $R=4 \, \mu m$, $\delta=76 \, nm$ and $\dot{\gamma} = 2.0 \, s^{-1}$. The light green curve is the best constant model that describes the data and the purple curve is the best sinusoidal model that describes the data.}
\label{fig:bad_fit}
\end{figure*}

\begin{figure}[h]
\centering
 \includegraphics[width=9cm]{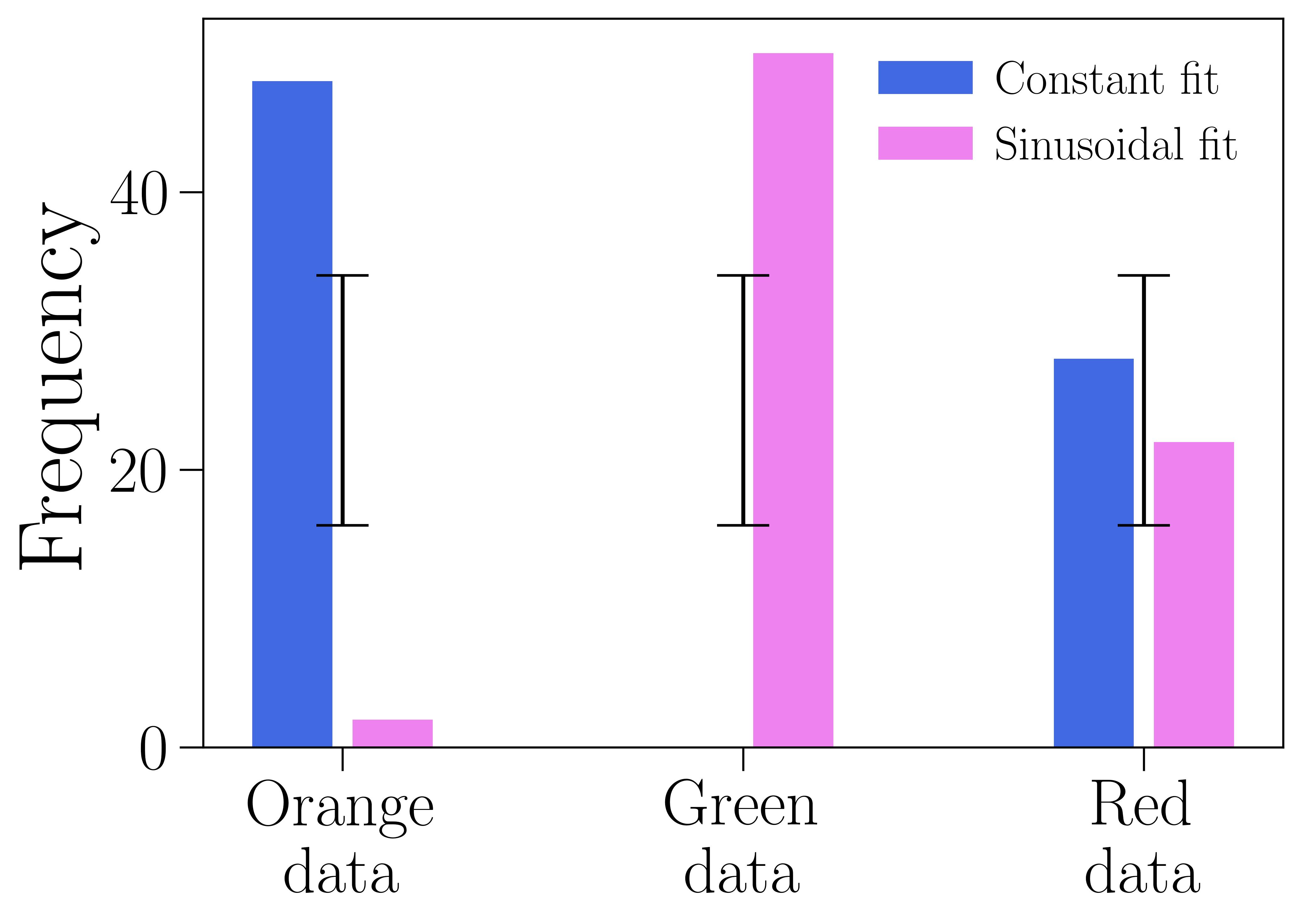}
 \caption{Outcome of classification of the state points examplified by the trajectories in Figure~\ref{fig:fit_traject}. The blue (pink) bar represents the number of particles classified as constant (sinusoidal). Black error bars represent the confidence interval of the null hypothesis.
 }
\label{fig:classif}
\end{figure}

Figure~\ref{fig:fit_traject} shows three examples of $\omega(\bar{t})$. Note that the curves
in the top panel are the same data shown in Figure~5 of the manuscript with matching colors (see caption). For each state point, all $\omega(\bar{t})$ are fitted to a constant and a sinusoidal model and the fit is compared using the Akaike Information Criterion,\cite{burnham2002model} i.e.,
\begin{equation}
    AIK = 2 k + n  \log(ssr / n) \, ,
 \label{eqn.7}
\end{equation}
where $k$ is the number of parameters in each model (one for constant, four for sinusoidal),
$n$ is the number of data points (which in the present context correspond to the number of distinct values of $\omega(\bar{t})$, i.e., the number of bins) and $ssr$ is the sum of the
squared difference between the data and the model (known as sum of squared residuals, hence $ssr$).
The model with the lowest $AIK$ is considered to be the best model for a trajectory.
Referring to Figure ~\ref{fig:fit_traject}, it is clear that the orange curve is best fitted by a constant model and, analogously, that the green
curve is best fitted by a sinusoidal model. Example results for fits to the red curve are shown
in Figure~\ref{fig:bad_fit}.
Clearly, both the constant and sinusoidal models are bad fits.
Nevertheless, the sinusoidal model turns out to be better because the peak associated with the only observed rotation introduces a very strong penalty for the constant model, which not only does not capture the peak well, but is also shifted upward by it and hence does not capture the constant part either.
The sinusoidal model, instead, is able to better ``absorb" the variations in the data and therefore turns out to be slightly better. Other realizations of the same state point, however, do not show the rotation and hence do not have a peak in $\omega(\bar{t})$, like the cyan trajectory shown in Figure~5B of the manuscript.
Those trajectories are better fitted by the constant model.
As this protocol is applied to each of the 50 particles simulated, it results in a binary classification of the form shown in Figure~\ref{fig:classif}.

Figure~\ref{fig:classif} shows the number of particles classified as constant or sinusoidal
for the three examples of Figure \ref{fig:fit_traject}.
A two-sided binomial test is applied to decide whether or not the classification is reliable.
To this aim, a null hypothesis is made that the data come from the random sampling of a binary variable, i.e., that they come from $B(k, 50, 0.5)$, the binomial distribution with 50 trials
(the number of particles simulated per state point) and success probability 0.5.
The threshold of the $p$ value is fixed at 0.01 and the confidence interval of the null hypothesis is computed.
If the frequency of both classes falls inside such an interval, the null hypothesis cannot be rejected, the state point is not classified and is marked as a red state point.
In the case where the null hypothesis can be rejected and the test result is assumed to be reliable, the state point is asigned to the more populated class (constant or sinusoidal).

\section {Dynamical behavior of the charged particle near a charged wall} \label{Sec:S3}

\begin{figure}[h]
\centering
 \includegraphics[width=9cm]{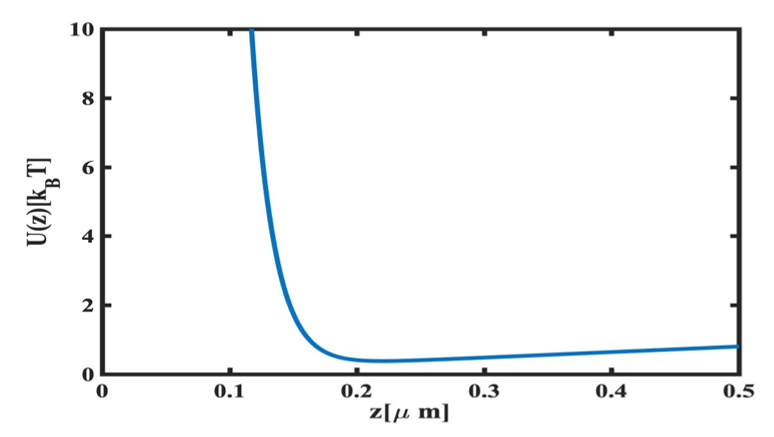}
 \caption{Interaction potentials for a negatively-charged polystyrene particle bounded by a negatively charged glass wall from equation (\ref{eqn.3}).}
\label{figS1}
\end{figure}

It is essential to verify the accuracy of the governing equations introduced in Section \ref{Sec:S1} for the system of interest, i.e., for a negatively charged particle interacting with a negatively charged wall. Therefore, the modeled electrostatic double-layer interaction potential, $U_{EDL}$, derived in Section \ref{Sec:S1} is compared to published Total Internal Reflection Microscopy (TIRM) data.
The TIRM experimental measurement shown in Figure 5A of Volpe et al.\cite{Volpe:09} was obtained using a polystyrene particle with radius $R = 1.45\,\mu m$ near a glass surface, density of $\rho_p = 1.053\,g/cm^3$, suspended in $300\,\mu M$ aqueous NaCl background electrolyte, and a Debye screening length $\kappa^{-1} = 17\,nm$ for the $1$:$1$ electrolyte. Using the same parameters in equation~(\ref{eqn.3}), the potential-distance relationship in Figure \ref{figS1} is obtained and shows good agreement with Volpe et al.'s\cite{Volpe:09} TIRM data.

\begin{figure}[h]
\centering
 \includegraphics[width=0.5\textwidth]{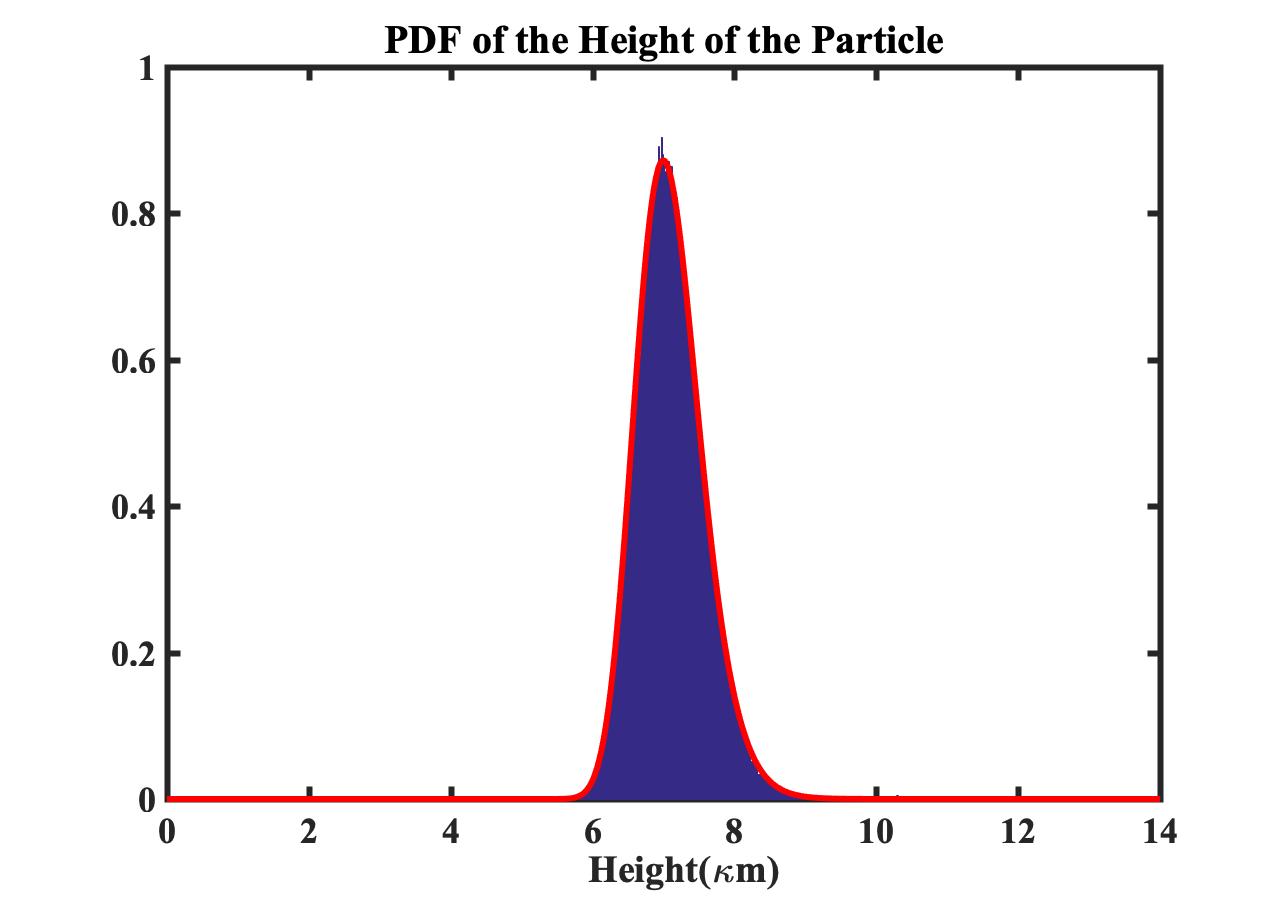}
 \caption[]{Comparison of the simulated PDF of the height of a SiO\textsubscript{2} particle of R = $1\,\mu m$ above a negatively charged wall (blue histogram) and the theoretically prediction from Boltzmann distribution (red curve).}
\label{figS2}
\end{figure}

For a spherical particle with a total energy of $U_{tot}(z)=U_{EDL}+U_{gravity}$ in thermal equilibrium with the surrounding fluid that obeys the one-dimensional stochastic differential equation~(\ref{eqn.4}), the probability distribution function (PDF) of the height of the particle is given by the Boltzmann distribution:
\begin{equation}
	\rho_{s}(z) = A_p\cdot e^{[-\frac{U_{tot}(z)}{k_BT}]}\ \,,
 \label{eqn.8}
\end{equation}
with $A_p$ chosen such that \(\int{p_s(z)}dz\)=1. Hence, in another check of the validity of the technique and the methodology, Figure \ref{figS2} shows the numerically calculated PDF (blue bars) of a SiO\textsubscript{2} particle with R = $1\,\mu m$ suspended in DI water (Millipore, resistivity $18.2$ M${\Omega}$cm and viscosity $\mu=1e^{-3} Pa\,s$ at $25$ °C, $I=1e^{-6} \,M$) at an equilibrium state overlaid with the theoretical Boltzmann distribution (red curve).
\

In order to describe the trajectory of a uniformly charged particle of radius $R$ suspended in a fluid and bounded by a wall, the electrostatic double later interaction $F_{EDL}$, equation (\ref{eqn.9}), as well as the effective gravitational force $F_g$, equation~(\ref{eqn.10}), are considered as non-hydrodynamic forces in equation~(\ref{eqn.4}), (i.e., ${F}^{nh}_i={F}_{EDL}+ {F}_g^*$).
 \begin{equation}
         {F}_{EDL}=2 \pi R \varepsilon_0 \kappa \frac{2 \psi_p\psi_w e^{-\kappa z}+(\psi^2_p+\psi^2_w) e^{-2\kappa z}}{1-e^{-2\kappa z}}      \,,
	 \label{eqn.9}
\end{equation}
 \begin{equation}
         {F}_{g}= -m^*\,g
 \label{eqn.10}
\end{equation}
where $m^*$ is the effective mass of the particle, i.e., the density mismatch between the particle and the surrounding fluid is taken into account.
Substituting equations~(\ref{eqn.9}) and~(\ref{eqn.10}) into equation~(\ref{eqn.4}) for the $F_i^{nh}$ term in the $z$ direction and the Brownian force, $F_i^B$, with the Gaussian stochastic process, $\eta_z(t)$, discussed in Section \ref{Sec:S1}, the final Langevin equation is obtained:

\begin{equation}
	\gamma \frac{dz}{dt}=F_{EDL}(z,\psi_p,\psi_w)+F_g+\ell \eta_z(t)
    \label{eqn.11}
\end{equation}
\\

Equation~(\ref{eqn.11}) is rendered dimensionless by measuring the length scale in units of Debye screening length $\kappa^{-1}$, i.e., $\bar{z}=\kappa z$, and the time scale in units of $T_{0}=\frac{3 \pi \mu R}{k_b T\kappa^2}$, i.e., $\bar{t}=t/T_0$.

\begin{figure}[h!]
\centering
 \includegraphics[width=0.4\textwidth]{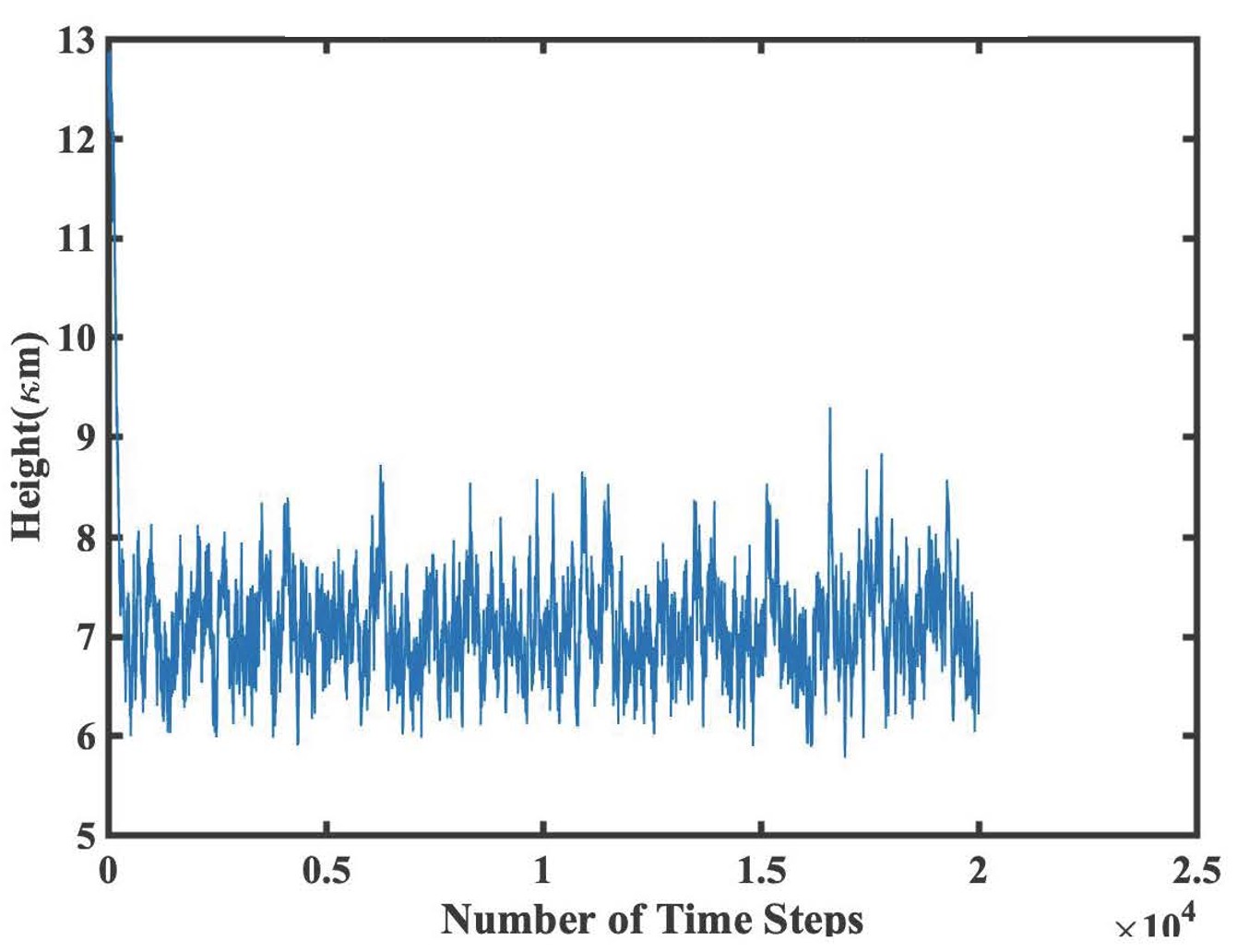}
 \caption{$1$D trajectory of a SiO\textsubscript{2} particle near a wall fluctuating around the equilibrium height that matches with the probability distribution function shown in Figure~\ref{figS2}.}
\label{figS3}
\end{figure}

Numeric integration of equation~(\ref{eqn.11}) utilizing the Euler time integration scheme yields the trajectories of the Brownian particle bounded by the wall. Figure \ref{figS3} displays the 1D height trajectory obtained for an SiO\textsubscript{2} particle of radius $R = 1\,\mu m$ bound by a negatively charged wall.

Agreement of the model with published experimental TIRM data (Fig. \ref{figS1}) and overlap of the simulated height distribution with theoretically predicted Boltzmann distribution (Fig. \ref{figS2}) instills confidence that the BD simulations can successfully simulate the dynamics of a bottom-heavy Janus particle near a wall under more complicated conditions.

\section {Impact of Shear Flow and Radius on Particle Behavior in Model III} \label{Sec:S4}

\begin{figure*}[!ht]
\centering
 \includegraphics[width=\textwidth]{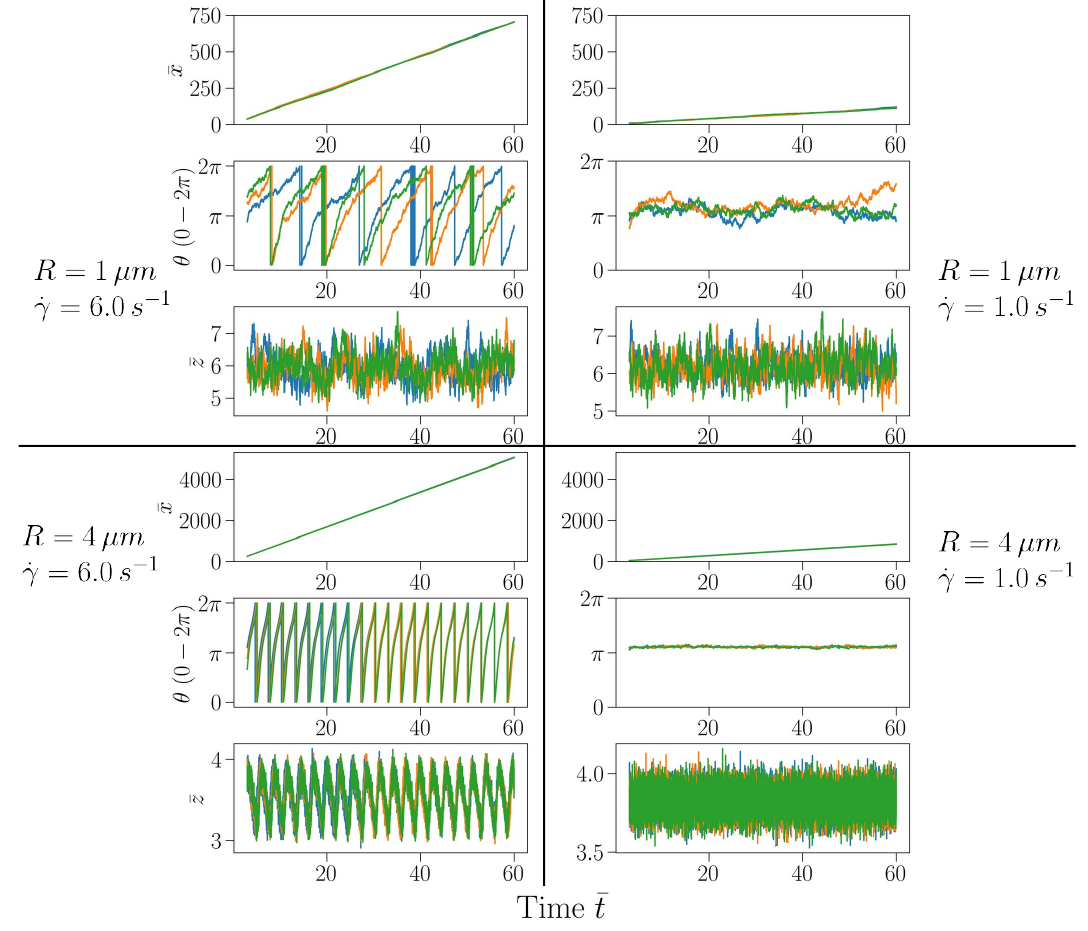}
 \caption[title]{Overall behavior of a particle in Model III.
 Surface charges are set to $\psi_1=-20\,mV$, $\psi_2=-40\,mV$ and $\psi_w=-50\,mV$, and $\delta=86\,nm$ for (top left quadrant) $R=1\,\mu m$ and $\dot{\gamma}=6.0\,s^{-1}$, (top right quadrant) $R=1\,\mu m$ and $\dot{\gamma}=1.0\,s^{-1}$, (bottom left quadrant) $R=4\,\mu m$ and $\dot{\gamma}=6.0\,s^{-1}$, and (bottom right quadrant) $R=4\,\mu m$ and $\dot{\gamma}=1.0\,s^{-1}$.
Three panels for each state from top to bottom are: non-dimensionalized particle distance from origin along x-axis $\bar{x}$, particle orientation $\theta$ and non-dimensionalized particle height $\bar{z}$ shown as a function of non-dimensionalized time $\bar{t}$.}
\label{fig:trajects_NoFrict}
\end{figure*}

The behavior of a particle using Model III is summarized in Figure \ref{fig:trajects_NoFrict} for four specific state points.
Surface charges are set to $\psi_1=-20\,mV$, $\psi_2=-40\,mV$ and $\psi_w=-50\,mV$, and $\delta=86\,nm$. Each state point is represented by three panels (from top to bottom): non-dimensionalized particle distance from origin along x-axis $\bar{x}$, particle orientation $\theta$ and non-dimensionalized particle height $\bar{z}$ shown as a function of non-dimensionalized time $\bar{t}$.
Figure \ref{fig:trajects_NoFrict} (top left quadrant) shows the state point for $R=1\,\mu m$ and $\dot{\gamma}=6.0\,s^{-1}$.
Figure \ref{fig:trajects_NoFrict} (top right quadrant) shows the state point for the same radius, $R=1\,\mu m$, but a decreased shear flow, $\dot{\gamma}=1.0\,s^{-1}$.
Figure \ref{fig:trajects_NoFrict} (bottom left quadrant) shows the state point for $R=4\,\mu m$ and $\dot{\gamma}=6.0\,s^{-1}$, while Figure \ref{fig:trajects_NoFrict} (bottom right quadrant) shows the state point for the same radius, $R=4\,\mu m$, but decreased shear flow, $\dot{\gamma}=1.0\,s^{-1}$.

Comparing panels in Figure \ref{fig:trajects_NoFrict} from left to right, it is clear that for smaller $\dot{\gamma}$, the particle travels less along the horizontal direction, the rotating
behavior disappears when the shear rate is too small and the height of the particle is highly correlated with its orientation, i.e., when the rotating behavior is observed, the height of the particle oscillates consistently with the orientation, while
in absence of rotation the height is also mainly determined by the noise of the Langevin
dynamics.

\section {Angular Velocity Maps and Particle Height for Model III}\label{Sec:S5}

\begin{figure*}[!ht]
\centering
 \includegraphics[width=\textwidth]{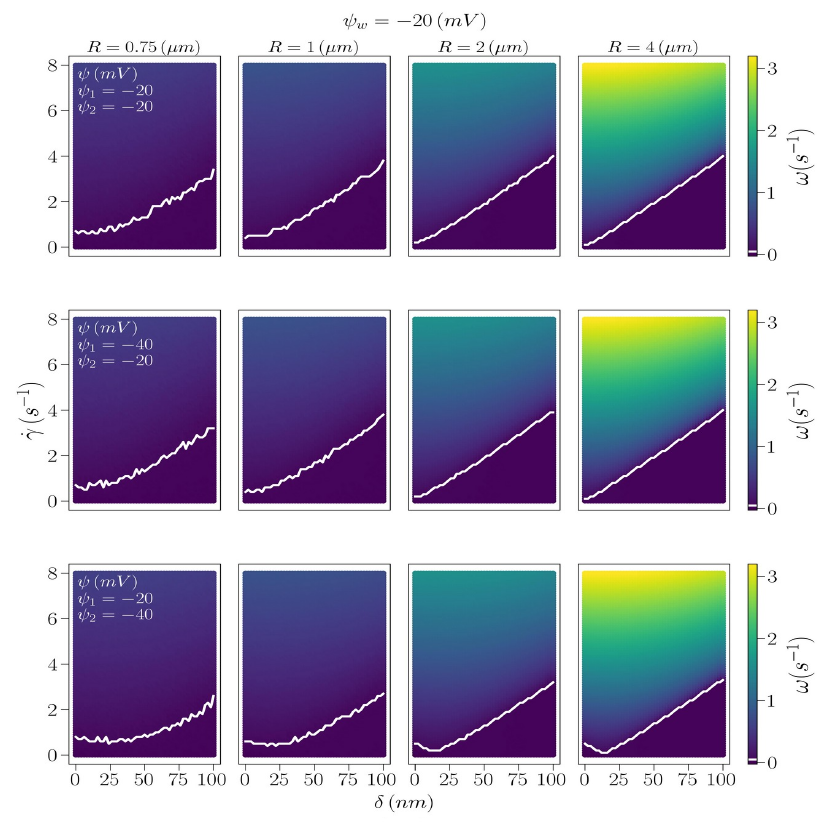}
 \caption[title]{Angular velocity maps of a bottom-heavy Janus particle under varying shear flow.
 The wall charge is set to $\psi_w=-20\,mV$. From top to bottom, particle surface charges are
 $\psi_1=-20\,mV$, $\psi_2=-20\,mV$(top),
 $\psi_1=-40\,mV$, $\psi_2=-20\,mV$(middle),
 $\psi_1=-20\,mV$, $\psi_2=-40\,mV$(bottom).
From left to right: $R=0.75, 1.00, 2.00,$ and $4.00 \, \mu m$.
 }
\label{fig:Omega_wall20}
\end{figure*}

\begin{figure*}[!ht]
\centering
 \includegraphics[width=\textwidth]{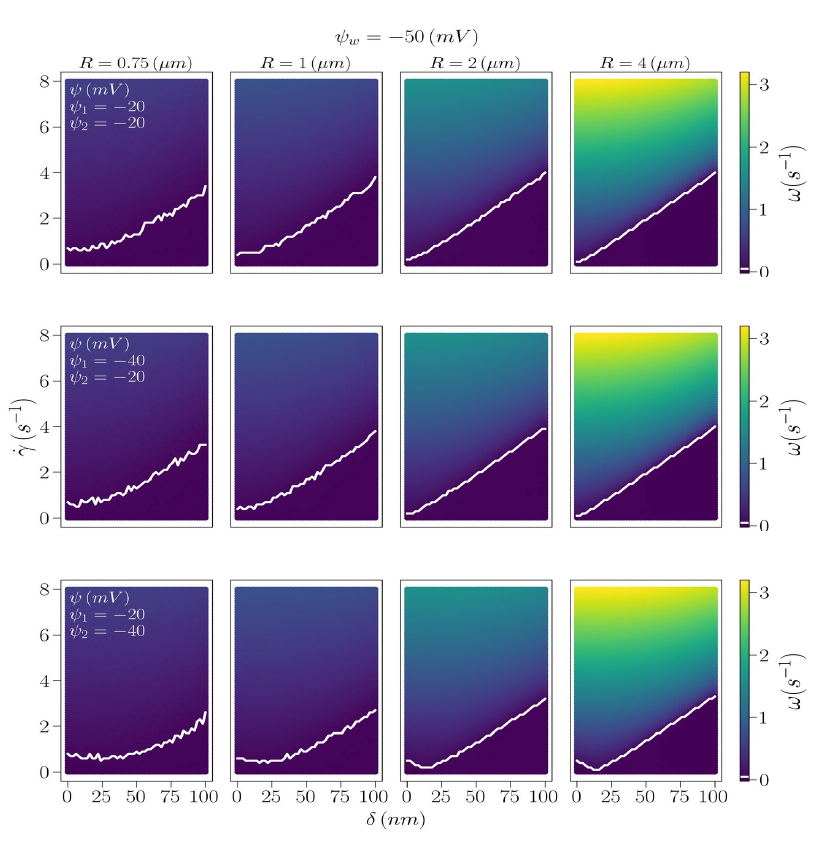}
 \caption[title]{Angular velocity maps of a bottom-heavy Janus particle under varying shear flow.
 The wall charge is set to $\psi_w=-50\,mV$. From top to bottom, particle surface charge are
 $\psi_1=-20\,mV$, $\psi_2=-20\,mV$(top),
 $\psi_1=-40\,mV$, $\psi_2=-20\,mV$(middle),
 $\psi_1=-20\,mV$, $\psi_2=-40\,mV$(bottom).
From left to right: $R=0.75, 1.00, 2.00,$ and $4.00 \, \mu m$.
 }
\label{fig:Omega_wall50}
\end{figure*}

Angular velocity maps, $\omega$-maps, are presented in this section as a function of $\delta$ and $\dot{\gamma}$ for a large class of systems.
Specifically, three combinations of surface charge of the particle are considered, i.e., $(\psi_1,\psi_2) = (-20, -20), \, (-20, -40), \, (-40, -20)$, all expressed in $mV$.
For each of the combinations, two values of the wall charge $\psi_w = -20\,mV$ and $\psi_w = -50\,mV$ are considered, hence specifying six classes of systems as they emerge from the electrostatic properties of the particle and the wall.
In each of the six classes, four particle radii $R=0.75, 1.00, 2.00,$ and $4.00 \, \mu m$ are considered.
The resulting $\omega$-maps are shown in Figures \ref{fig:Omega_wall20} and \ref{fig:Omega_wall50} for $\psi_w = -20\,mV$ and $\psi_w = -50\,mV$, respectively, with $R$ increasing from left to right.

Figures \ref{fig:Omega_wall20} and \ref{fig:Omega_wall50} further support the analysis provided in Section 6 of the manuscript: Systems with $(\psi_1=\psi_2) = (-20, -20), \, (-40, -20)$ are characterized by a monotonic dependence of the white transition line on $\dot{\gamma}$ as $\delta$ grows.
For the system with  $(\psi_1,\psi_2) = (-20, -40)$, instead, the minimum around $\delta=16 \, nm$ is systematically observed, regardless of $\psi_w$.
For larger particles, the non-monotonicity is particularly visible as the data become noisier when $R$ decreases.
This, in turn, is caused by the small range of $\omega$ values, which makes the transition line more noisy, not only in the proximity of the minimum.
Nonetheless, it can be seen that on reducing $R$ the minimum persists, but is less sharp.
Overall, the results across all systems confirm that the presence of the minimum is a consequence of the charge imbalance between the two faces of the Janus particle and, in particular, that it is observed only
when the heavy side of the particle is overcharged.

\begin{figure*}[!ht]
\centering
 \includegraphics[width=\textwidth]{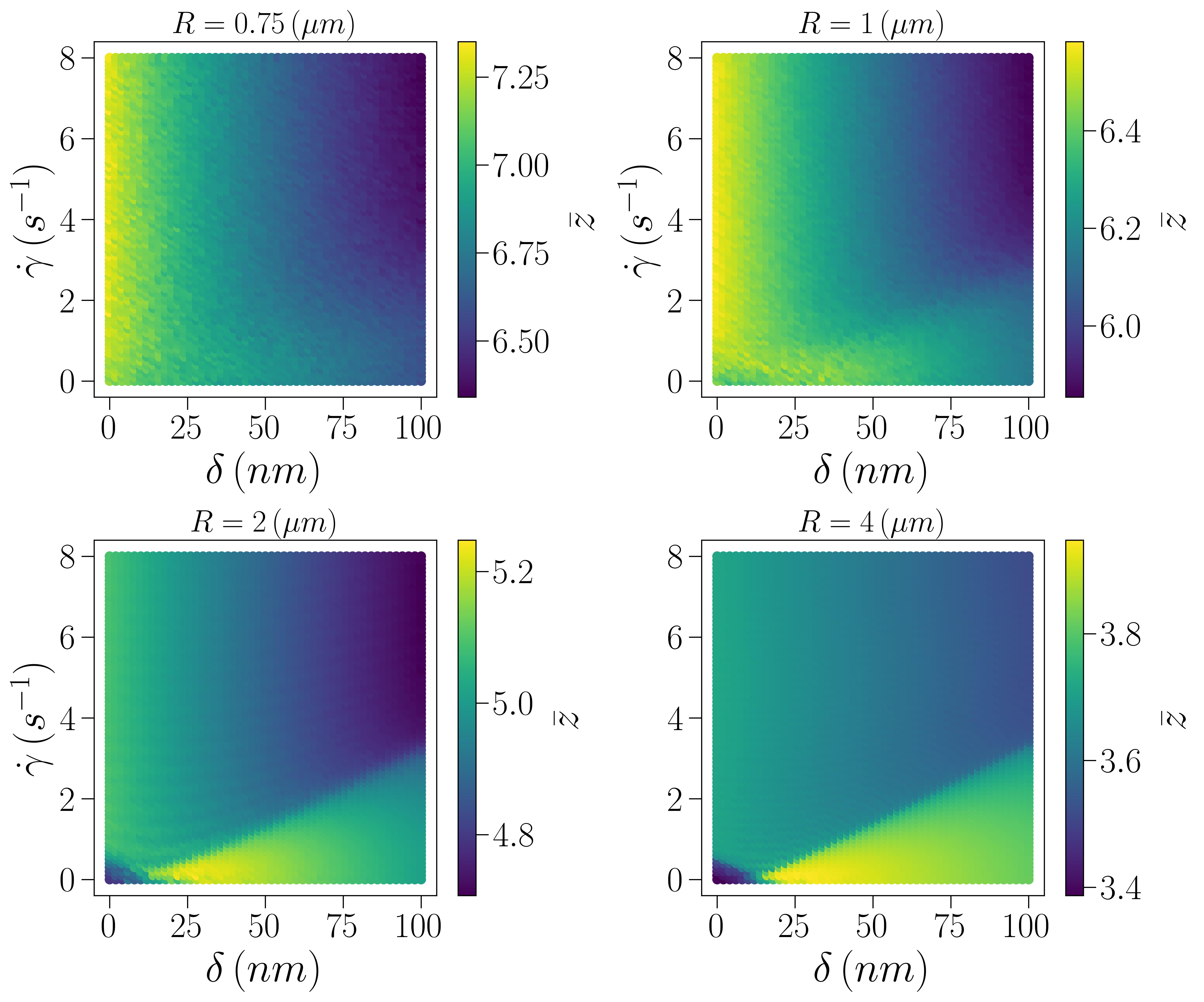}
 \caption[title]{Average non-dimensionalized height, $\bar{z}$, as a function of cap thickness $\delta$ and shear rate, $\dot{\gamma}$, for bottom-heavy particles with radius $R=0.75, 1, 2,$ and $4\,\mu m$.
Surface charges are set to $\psi_1=-20 mV$, $\psi_2=-40\,mV$ and $\psi_w=-50\,mV$.}
\label{fig:Zmaps}
\end{figure*}

Figure \ref{fig:Zmaps} shows the average non-dimensionalized height, $\bar{z}$, as a function of cap thickness, $\delta$, and shear rate, $\dot{\gamma}$, for bottom-heavy particles with radius $R=0.75, 1, 2,$ and $ 4\,\mu m$.
Surface charges are set to $\psi_1=-20\,mV$, $\psi_2=-40\,mV$ and $\psi_w=-50\,mV$.
The strong correlation between height and the behavior of $\theta$ seen in Figure ~\ref{fig:trajects_NoFrict} is also apparent in Figure~\ref{fig:Zmaps}, where the  non-rotation region exists for all particle sizes except the smallest one.
Note that particles with the smallest size considered in this study are also those that show the slowest rotation at given state points.

\section {State Diagrams for Model III} \label{Sec:S6}
\begin{figure*}[!ht]
\centering
 \includegraphics[width=\textwidth]{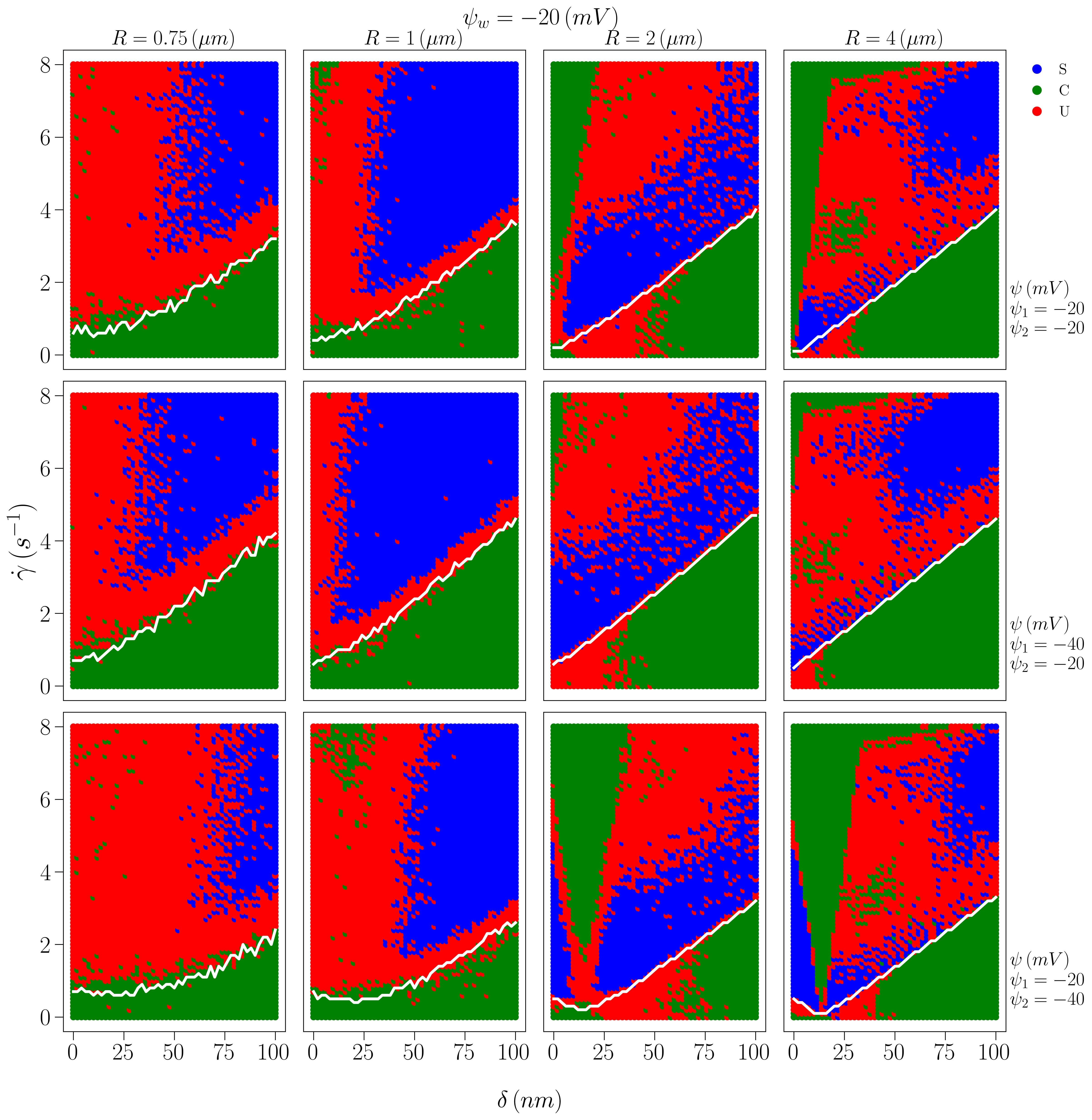}
 \caption[title]{State diagrams of a bottom-heavy Janus particle under varying shear flow.
 The wall charge is set to $\psi_w=-20\,mV$. From top to bottom, particle surface charge are
 $\psi_1=-20\,mV$, $\psi_2=-20\,mV$(top),
 $\psi_1=-40\,mV$, $\psi_2=-20\,mV$(middle),
 $\psi_1=-20\,mV$, $\psi_2=-40\,mV$(bottom).
From left to right: $R=0.75, 1.00, 2.00,$ and $ 4.00 \, \mu m$.
 }
\label{fig:SD_wall20}
\end{figure*}

\begin{figure*}[!ht]
\centering
 \includegraphics[width=\textwidth]{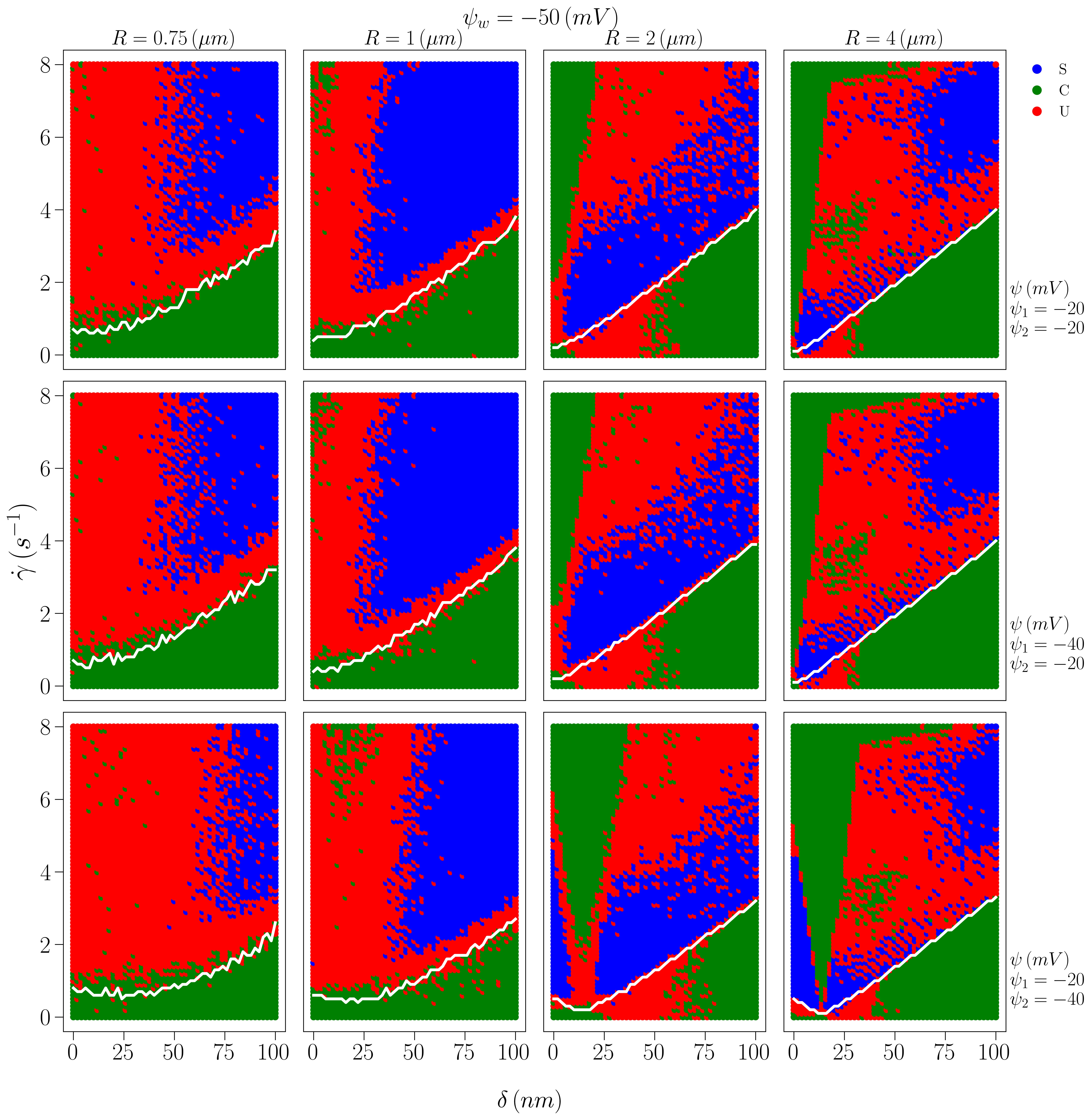}
 \caption[title]{State diagrams of a bottom-heavy Janus particle under varying shear flow.
 The wall charge is set to $\psi_w=-50\,mV$. From top to bottom, particle surface charge are
 $\psi_1=-20\,mV$, $\psi_2=-20\,mV$(top),
 $\psi_1=-40\,mV$, $\psi_2=-20\,mV$(middle),
 $\psi_1=-20\,mV$, $\psi_2=-40\,mV$(bottom).
From left to right: $R=0.75, 1.00, 2.00,$ and $ 4.00 \, \mu m$.
 }
\label{fig:SD_wall50}
\end{figure*}

State diagrams derived using the analysis described in Section \ref{Sec:S2} are shown in  Figures~\ref{fig:SD_wall20} and~\ref{fig:SD_wall50} for the same systems shown in Section \ref{Sec:S5}.
The transition line (white) between the rotating and non-rotating state is generally well-identified, with the entire non-rotating region classified as constant except for large particles with small cap thickness at small shear flow for which a relatively small region of unclassified state points is observed.
As such regions are in the vicinity of the transition line, they are ascribed to the peculiar behavior of the system as exemplified by Figure 5 in the manuscript.
In brief, at large $\delta$ and large $\dot{\gamma}$, there is a vast region of state points classified as sinusoidal, with the extension of the region weakly affected by the particle radius and affected by the charge imbalance.
The surface charge value seems indeed to be the more relevant parameter in determining the transition from rotating states with constant angular velocity and the sliding behavior with time-varying angular velocity.
Reducing $\delta$ implies to leave the region of state points classified as sinusoidal, i.e., the ones that are characterized by sliding behavior, and to enter a region of constant angular velocity.
Such region has a triangular shape and ''moves" in the diagram with changing parameters consistent with the interpretation given in the manuscript.

\section {Impact of Hydrodynamic Friction in Model IV}\label{Sec:S7}
\begin{figure*}[!ht]
\centering
 \includegraphics[width=\textwidth]{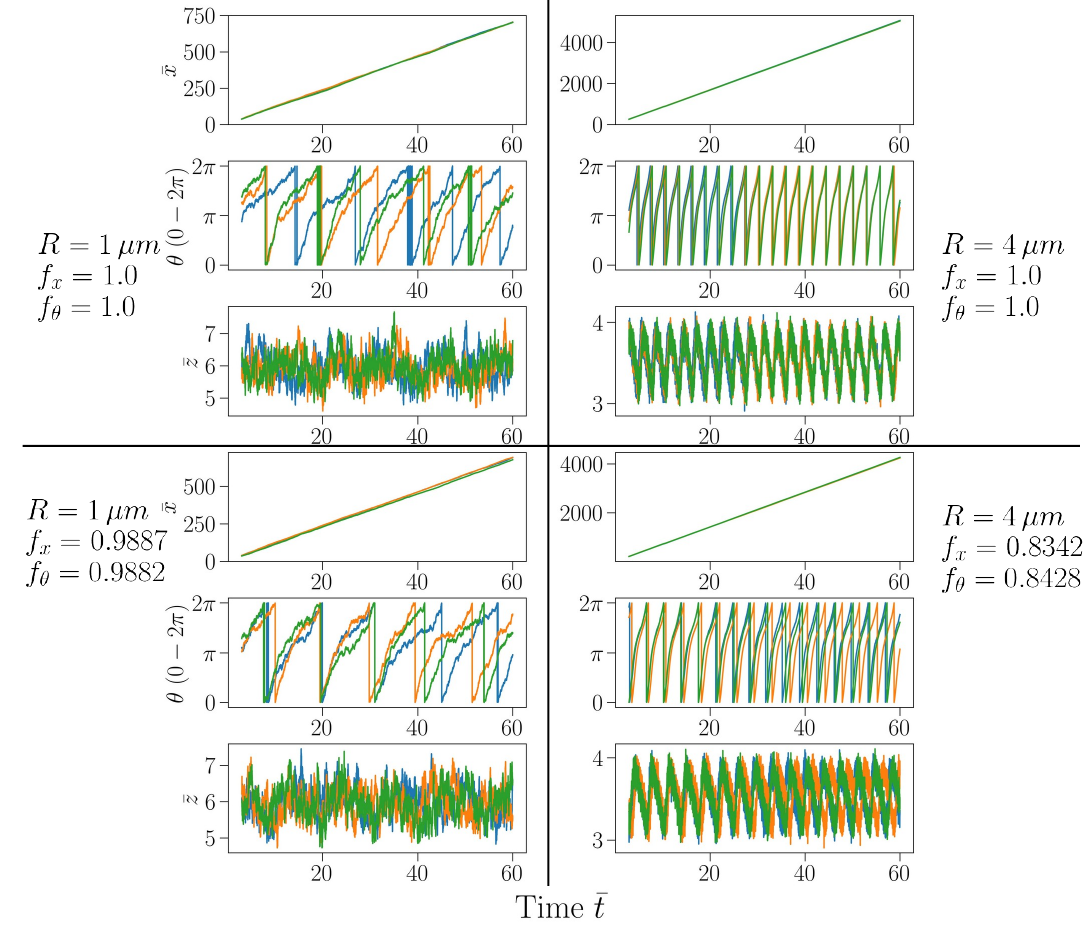}
 \caption[title]{Impact of hydrodynamic friction - overall behavior of a bottom-heavy Janus particle in Model III (top) vs Model IV (bottom) for particle with $R=1\,\mu m$ (left,  $f_x=f_{\theta}=1$ vs $f_x = 0.9887$ and $f_{\theta} = 0.9882$.) and $R=4$ $\mu m$ (right, $f_x=f_{\theta}=1$ vs. $f_x = 0.8342$ and $f_{\theta} = 0.8428$). Surface charges are set to $\psi_1=-20\,mV$, $\psi_2=-40\,mV$ and $\psi_w=-50\,mV$, and $\delta=86\,nm$ and $\dot{\gamma}=6.0\,s^{-1}$.
Three panels for each state from top to bottom are: non-dimensionalized particle distance from origin along x-axis $\bar{x}$, particle orientation $\theta$ and non-dimensionalized particle height $\bar{z}$ shown as a function of non-dimensionalized time $\bar{t}$.}
\label{fig:HDF_20-40-50_R=1_-w-wo}
\end{figure*}

The dynamical behavior of a bottom-heavy Janus particle using Models III and IV without (top two quadrants) and with friction (bottom two quadrants), respectively, is compared in Figure~\ref{fig:HDF_20-40-50_R=1_-w-wo} for two particle sizes, $R=1$ (left) and $R=4\,\mu m$ (right).
Each state point is represented by three panels (from top to bottom): non-dimensionalized particle distance from origin along x-axis $\bar{x}$, particle orientation $\theta$ and non-dimensionalized particle height $\bar{z}$ shown as a function of non-dimensionalized time $\bar{t}$.
Surface charges are set to $\psi_1=-20\,mV$, $\psi_2=-40\,mV$ and $\psi_w=-50\,mV$, while cap size is $\delta=86\,nm$ and the shear rate is $\dot{\gamma}=6.0\,s^{-1}$.
Friction factors are $f_x=f_{\theta}=1$ for the top two quadrants in the absence of hydrodynamic friction. In the presence of hydrodynamic friction, friction factors based on Goldman et al.\cite{goldman1967slow} of $f_x = 0.9887$ and $f_{\theta} = 0.9882$ are used for $R=1$ $\mu m$ (bottom left quadrant) and $f_x=0.8342$, $f_{\theta}=0.8428$ for $R=4$ $\mu m$ (bottom right quadrant).

Comparing each top panel with the corresponding lower panel shows that hydrodynamic friction has virtually no impact on the $R=1\,\mu m$ particle behavior beyond reducing the maximum distance traveled, which diminishes when hydrodynamic friction is accounted for.
In the $R=4\,\mu m$ particle case, the particle also travels less far, but additionally experiences fewer rotations due to hydrodynamic friction, which is expected since it is closer to the wall due to its larger size.

\clearpage


\providecommand*{\mcitethebibliography}{\thebibliography}
\csname @ifundefined\endcsname{endmcitethebibliography}
{\let\endmcitethebibliography\endthebibliography}{}

\end{document}